\documentclass[10pt, journal, final]{IEEEtran}
\IEEEoverridecommandlockouts
\usepackage{bbm}
\usepackage{amsmath}
\usepackage{amsfonts}
\usepackage[dvips]{graphicx}
\usepackage{times}
\usepackage{cite}
\usepackage{array}
\usepackage{mathrsfs}
\usepackage{multirow}
\usepackage{amssymb}

\usepackage{stfloats}
\usepackage{subfigure}
\usepackage{footnote}
\usepackage{booktabs}
\usepackage{array}
\usepackage{algorithm}
\usepackage{subeqnarray}
\usepackage{cases}
\usepackage{threeparttable}
\usepackage{xcolor}
\usepackage{hyperref}
\usepackage{epstopdf}
\usepackage{algpseudocode}
\usepackage{bm}
\usepackage{multirow}
\usepackage{adjustbox}
\usepackage{stfloats}
\usepackage{diagbox}

\def\BibTeX{{\rm B\kern-.05em{\sc i\kern-.025em b}\kern-.08em
		T\kern-.1667em\lower.7ex\hbox{E}\kern-.125emX}}

\begin{document}
	
\title{Hierarchical Beam Alignment for Millimeter-Wave Communication Systems: A Deep Learning Approach}


\author{
Junyi~Yang, Weifeng~Zhu, Meixia~Tao, and Shu~Sun
\thanks{
This paper was presented in part at the IEEE Global Conference of Communications (GLOBECOM) 2022 \cite{Globecom2022}}
\thanks{The authors are with the Department of Electronic Engineering, Shanghai Jiao Tong University, Shanghai 200240,
China (e-mail: \{yangjunyi, wf.zhu, mxtao, shusun\}@sjtu.edu.cn).}

}
\maketitle	
\vspace{-1cm}

\begin{abstract}
Fast and precise beam alignment is crucial for high-quality data transmission in millimeter-wave (mmWave) communication systems, where large-scale antenna arrays are utilized to overcome the severe propagation loss. To tackle the challenging problem, we propose a novel deep learning-based hierarchical beam alignment method for both multiple-input single-output (MISO) and multiple-input multiple-output (MIMO) systems, which learns two tiers of probing codebooks (PCs) and uses their measurements to predict the optimal beam in a coarse-to-fine search manner. Specifically, a hierarchical beam alignment network (HBAN) is developed for MISO systems, which first performs coarse channel measurement using a tier-1 PC, then selects a tier-2 PC for fine channel measurement, and finally predicts the optimal beam based on both coarse and fine measurements. The propounded HBAN is trained in two steps: the tier-1 PC and the tier-2 PC selector are first trained jointly, followed by the joint training of all the tier-2 PCs and beam predictors. Furthermore, an HBAN for MIMO systems is proposed to directly predict the optimal beam pair without performing beam alignment individually at the transmitter and receiver. Numerical results demonstrate that the proposed HBANs are superior to the state-of-art methods in both alignment accuracy and signaling overhead reduction.
\end{abstract}


\section{Introduction}

Millimeter-wave (mmWave) communication is expected to play a key role in the fifth generation (5G) and beyond cellular networks for its abundant spectrum resource located between about 28 GHz and 300 GHz \cite{overview1,overview2,overview3}.
Compared with the conventional sub-6 GHz counterpart, mmWave communication provides significantly larger bandwidth and orders of magnitude higher bit-rates. Meanwhile, it also usually suffers from harsher propagation environments including severe free-space path loss and poor diffraction. 
Fortunately, the short wavelength at mmWave bands supports more antennas packed in the same physical space, which allows sharper directional beamforming with large-scale antenna arrays to compensate for the path loss \cite{compensate}. 
Due to the undesirable cost and power consumption of an all-digital beamforming architecture at mmWave bands, the analog-only or hybrid analog/digital architecture is usually implemented in mmWave communication systems\cite{hybrid1}\cite{hybrid2}.
There are two broad categories of beamforming methods: non-codebook-based beamforming and codebook-based beamforming. Non-codebook-based beamforming usually refers to designing the optimal beam under a total power constraint only, while codebook-based beamforming aims to search the best beam from a pre-defined codebook thus is also known as beam alignment. Compared with non-codebook-based beamforming, beam alignment usually requires less channel feedback and computation overhead. Therefore, we focus on beam alignment in this work. However, the large-scale antenna array results in massive narrow beams in the codebook to be measured, and mmWave communication is very sensitive to the dynamic environment. Therefore, it is crucial to design a high-accuracy and low-overhead beam alignment method for mmWave systems.

\subsection{Related Works}

Beam alignment methods usually employ beam sweeping to measure the channel conditions between each base station (BS) and each user equipment (UE), and then select the optimal directional beamformer from the codebook based on the measurements \cite{sweeping}. In the common method known as exhaustive search, the BS and the UE sweep all the beamformer pairs in the codebook to choose the one with maximal received power. However, such a brute-force sweeping method suffers from prohibitive signaling overhead and undesirable latency in the system with large-scale antenna arrays, which limits its usage in practical mmWave systems \cite{exhaustive1, exhaustive2}.
As an alternative approach, hierarchical beam search can greatly reduce the signaling overhead and alignment latency by utilizing a pre-designed multiple-tier codebook \cite{hierarchical1,hierarchical2,hierarchical_design2,hierarchical_design1}. 
Specifically, the BS usually sweeps several wider beams and then gradually focuses on a thinner search space to seek the optimal beam. 
Note that a well-designed hierarchical codebook can effectively improve the performance of the optimal beam search \cite{hierarchical_design1,hierarchical3}. In \cite{hierarchical_design1}, a complete binary-tree structured hierarchical codebook is designed by jointly utilizing sub-array and deactivation techniques. 
In \cite{hierarchical3}, an alternating minimization method with a closed-form expression (AMCF) for wide beam design in hierarchical codebooks is proposed, where the codewords are formed without antenna de-activation.
The hierarchical beam search frameworks are also designed for the reconfigurable intelligent surface assisted systems \cite{hierarchical_design3,hierarchical_design4}. 
However, the performance of the hierarchical search method is sensitive to the wide beams with imperfect patterns and noise, which can easily result in error accumulation in the search process and limit the number of tiers for hierarchical codebook design.
In contrast to the above two methods, compressive beam alignment methods can also realize accurate beam alignment with a few measurements by adopting the advanced compressive sensing (CS) algorithms \cite{CS1, CS2}. However, CS algorithms usually suffer from high computational complexity when the antenna number is large.

Recently, deep learning (DL) is regarded as a promising technology to improve both the accuracy and the signaling overhead of beam alignment. 
In \cite{context1,context2,context3,context4}, deep neural networks (DNNs) are designed to perform beam prediction based on the context information of the UE location without beam sweeping. However, acquiring the location information of UEs needs additional sensors and additional feedback. Moreover, the context information can only provide limited performance improvement since the dynamic channel environment cannot be exploited.
On the other hand, deep reinforcement learning (DRL) approaches are also proposed to gradually find the optimal beam by performing repeated interactions between the BS and the UE \cite{reinforcement1,reinforcement2,reinforcement3}.
However, the frequent interactions in DRL may lead to higher latency and more control overhead, which limits their practical applications.
To alleviate the signaling overhead, the work \cite{beamforming1} proposes adopting the DNN to accurately find the optimal beam by directly sweeping a subset of codeword candidates in the pre-defined codebook.
To effectively capture the characteristics of the propagation environment and the channel information, the work \cite{Jeffrey} proposes to additionally learn a probing codebook (PC) along with a beam predictor. 
In \cite{beamforming2}, the beam patterns are also learned to better adapt to the surrounding environment under the DL framework.

Although these DL-based methods (e.g., \cite{beamforming1,Jeffrey}) can provide accurate beam alignment performance with reduced measurements, their performance improvement is limited if the size of the PC is not large. To obtain better measurement of the channel information of each UE, we are inspired to perform beam alignment by developing a DL-based hierarchical search framework. Generally, the channel information of each UE is firstly coarsely measured and then the optimal beam is predicted with additional fine search measurements. It is also noted that the aforementioned works \cite{reinforcement1,reinforcement2,reinforcement3,beamforming1,Jeffrey} all consider separately finding the optimal beams at the BS and the UE if both of them are equipped with multiple antennas. This work makes progress to take advantage of the mmWave channel structure to perform BS-UE beam pair alignment, which can further reduce the measurement overhead. 
In addition, since beam sweeping, measurement, and reporting are essential for current releases of 5G \cite{5G_framework}, the proposed beam sweeping based method can also fit into the 5G standard.
Concurrent to our work, the idea of directly synthesizing the transmit and receive beams are also studied in \cite{Jeffrey_arxiv}.


\subsection{Main Contributions}
 
This paper considers the beam alignment problem in mmWave communication systems. To effectively probe comprehensive channel information, we propose a novel hierarchical beam alignment framework applicable to both multiple-input single-output (MISO) and multiple-input multiple-output (MIMO) systems, which leverages DL techniques to seek the optimal directional beamformer with learnable PCs.
The main contributions can be summarized as follows.

\textbf{DL-Based hierarchical search framework:} We first propose a DL-based hierarchical search method for beam alignment in MISO systems. Specifically, we employ two tiers of learnable PCs and propose a hierarchical beam alignment network (HBAN) for beam alignment in a coarse-to-fine search manner. 
Similar to \cite{Jeffrey}, the one-layer complex neural networks (NNs) are used to model PCs. Particularly, the application of the hierarchical structure enriches the PC space for significant performance improvement. We also introduce an NN-based selector for coarse search and multiple NN-based beam predictors for fine search in the HBAN.

\textbf{Hierarchical beam pair alignment:} For MIMO systems where both the BS and the UE are equipped with multiple antennas, we propose to directly find the optimal BS-UE beam pair under a hierarchical search framework. 
Specifically, we develop an HBAN-MIMO for joint beam alignment at both BS and UE, where multiple learnable beam pairs are included as the codewords in each PC. Benefiting from the powerful DL technique, the proposed BS-UE beam pair alignment method can provide dramatic improvement in beam alignment accuracy and measurement cost compared with the conventional two-stage beam alignment methods.

\textbf{Two-step NN training strategy:} By considering the hierarchical search framework, we propose an effective training strategy in which the two tiers of the network are trained sequentially to avoid overfitting. In the first step, we propose a clustering-based method to generate labels, then train the tier-1 PC and the selector jointly for coarse channel measurement. After completing the training in the first step, we train the tier-2 PCs and beam predictors jointly for fine channel measurements. Both HBAN-MISO and HBAN-MIMO are trained based on the two-step training strategy.

\textbf{Numerical validation in various propagation environments:} We simulate the proposed method leveraging ray-tracing datasets of outdoor and indoor environments generated by a widely-used ray-tracing software Wireless InSite \cite{software}.
In both MISO and MIMO systems, our method can consistently achieve a significantly higher beam alignment accuracy compared to other DL-based methods with the same sweeping overhead. In addition, we examine the spectral efficiency and robustness of all considered methods and the numerical results show the superior performance of our method compared to other existing alternatives.

\subsection{Paper Organization and Notations}
The rest of this paper is organized as follows. The system model and problem formulation are described in Section \uppercase\expandafter{\romannumeral2}. Section \uppercase\expandafter{\romannumeral3}  introduces the proposed beam alignment method when only the BS is supposed to perform beam alignment while Section \uppercase\expandafter{\romannumeral4} explains the joint beam alignment framework for the case that both the BS and the UE are equipped with antenna arrays.
We show the simulation results in Section \uppercase\expandafter{\romannumeral5} and finally we conclude this paper in \uppercase\expandafter{\romannumeral6}.

In this paper, upper-case and lower-case letters denote random variables and their realizations, respectively. Boldface lower-case letters,boldface upper-case letters and calligraphy letters such as $\mathbf{x}$,$\mathbf{X}$ and $\mathcal{X}$ denote vectors, matrices and sets, respectively. Superscripts $(\cdot)^T$,$(\cdot)^*$ and  $(\cdot)^H$ denote transpose, conjugate and conjugate transpose, respectively. Further, $|\cdot|$ denotes the magnitude of a complex variable or a vector, depending on the context.

\section{System Model and Problem Formulation}
We consider a narrowband mmWave communication system with one BS and one UE, where the BS is equipped with a large number $M_{\rm{t}} \gg 1$ of antennas and the UE is equipped with $M_{\rm{r}} \ge 1$ antennas. The narrowband mmWave channel can be described by a multipath model \cite{analog_beamformer}. 
For simplicity, we assume that the antenna arrays are uniform linear arrays (ULAs) while our proposed method can be extended to other array geometries and 3D channel models. 
We only consider the radio frequency (RF) domain analog beamforming for the purpose of beam alignment, where each antenna element is connected to a dedicated analog phase shifter. The analog beamformer is given as \cite{analog_beamformer}
\begin{equation}\label{equ:beamformer}
\mathbf{v} = \frac{1}{\sqrt{M}}\left[e^{\mathrm{j}\phi_{1}},e^{\mathrm{j}\phi_{2}} ,\dots,e^{\mathrm{j}\phi_{M}}\right]^T,
\end{equation}
where $\phi_i$ is the phase of the $i$th element and $M$ represents the number of antennas in the array.

The commonly used discrete Fourier transform (DFT) codebook $\mathbf{V} = [\mathbf{v}_1, \mathbf{v}_2, \dots,\mathbf{v}_{N_{\rm{t}}}] \in \mathbb{C}^{M_{\rm{t}} \times N_{\rm{t}}}$ is adopted at the BS for data transmission in this work. 
In the DFT codebook, each beam steers to a discrete direction and the beam codeword $\mathbf{v}_{i}$ can be expressed as
\begin{align}\label{equ:DFT_beamformer}
    \mathbf{v}_{i} &= \frac{1}{\sqrt{M_{\rm{t}}}} \left[ 1,e^{\mathrm{j}\omega_i},\dots,e^{\mathrm{j}(M_{\rm{t}}-1)\omega_i} \right]^T, \notag \\ 
    \quad &\quad\quad\quad\quad\quad\quad\quad\quad\quad\quad i \in \{ 1,2,\dots,N_{\rm{t}} \},
\end{align}
where $\omega_i =\frac{2\pi d}{\lambda}\frac{(2(i-1)-N_{\rm{t}})}{N_{\rm{t}}}$ with $\lambda$ and $d$ denoting the carrier wavelength and antenna spacing, respectively. The codebook size $N_{\rm{t}}$ is usually assumed to be large enough to cover the whole space. 
Depending on whether $M_{\rm{r}} = 1$ or $M_{\rm{r}} > 1$, the mmWave communication system can be regarded as the MISO system or the MIMO system.
In the following subsections, we introduce the signal model and the problem formulation in these two systems.


\subsection{MISO System}
For the MISO system with $M_{\rm{r}} = 1$, we only need to perform beam alignment at the BS, where one of the beam in the pre-defined DFT codebook $\mathbf{V}$ is sought to realize the maximal signal-to-noise ratio (SNR) for transmission. When a symbol $s \in \mathbb{C}$ with unit power constraint $\mathbb{E}(|s|^2) = 1$ is transmitted using beam $\mathbf{v}_i$, the received signal $y$ at the UE can be expressed as
\begin{equation}\label{equ:single_y}
y = \sqrt{\rho}\mathbf{h}^H\mathbf{v}_is + n,
\end{equation}
where $\mathbf{h}\in\mathbb{C}^{M_{\rm{t}}\times1}$ is the narrowband mmWave channel between the BS and the UE ,$\rho$ is the transmit power and $n$ is the complex Gaussian noise with zero mean and variance $\sigma^2_{n}$. The received SNR at the UE with channel $\mathbf{h}$ using beam $\mathbf{v}_i$ can be written as
\begin{equation}\label{equ:single_SNR}
\text{SNR} = \frac{\rho|\mathbf{h}^H \mathbf{v}_i|^2}{\sigma^2_{n}}.
\end{equation}
For the given DFT codebook $\mathbf{V}$ consisting of $N_{\rm{t}}$ beams, our target is to find the optimal beam codeword that realizes the maximal SNR for transmission: 
\begin{align}\label{equ:single_problem}
i^* 
&=\mathop{\arg\max}\limits_{i \in \{ 1,2,\dots,N_{\rm{t}} \} } \left(\frac{\rho|\mathbf{h}^H\mathbf{v}_i|^2}{\sigma^2_{n}}\right) \notag \\
&= \mathop{\arg\max}\limits_{i \in \{ 1,2,\dots,N_{\rm{t}} \} }(|\mathbf{h}^H\mathbf{v}_i|^2).
\end{align}

\subsection{MIMO System}

For MIMO systems with $M_{\rm{r}} > 1$, we are supposed to perform beam alignment at both the BS and the UE. 
Similar to the BS, the UE is also equipped with a pre-defined DFT codebook $\mathbf{W} = [\mathbf{w}_1, \mathbf{w}_2, \dots,\mathbf{w}_{N_{\rm{r}}}] \in \mathbb{C}^{M_{\rm{r}} \times N_{\rm{r}}}$, where $N_{\rm{r}}$ represents the size of the DFT codebook. 
When a symbol $s \in \mathbb{C}$ with unit power constraint $\mathbb{E}(|s|^2) = 1$ is transmitted with beamformer $\mathbf{v}_i$ at the BS and $\mathbf{w}_j$ at the UE, the received signal $y$ can be expressed as 
\begin{equation}\label{equ:both_y}
y = \sqrt{\rho}\mathbf{w}_j^H\mathbf{H}^H\mathbf{v}_is + n,
\end{equation}
where $\mathbf{H} \in \mathbb{C}^{M_{\rm{t}} \times M_{\rm{r}}} $ is the narrowband MIMO mmWave channel between the BS and the UE. Then the received SNR at the UE can be written as

\begin{equation}\label{equ:both_SNR}
\text{SNR} = \frac{\rho|\mathbf{w}_j^H\mathbf{H}^H \mathbf{v}_i|^2}{\sigma^2_{n}}.
\end{equation}
The target is then to find the optimal beam pair from pre-defined DFT codebooks $\mathbf{V}$ and $\mathbf{W}$ to realize the maximal SNR transmission, which can be formulated as 
\begin{align}\label{equ:both_problem}
\{ i^*,j^* \} 
&=
\mathop{\arg\max}\limits_{ \substack{i \in \{ 1,2,\dots,N_{\rm{t}} \} \\ j \in \{ 1,2,\dots,N_{\rm{r}} \}} }  \left(\frac{\rho|\mathbf{w}_j^H\mathbf{H}^H \mathbf{v}_i|^2}{\sigma^2_{n}}\right) \notag \\
&= 
\mathop{\arg\max}\limits_{\substack{i \in \{ 1,2,\dots,N_{\rm{t}} \} \\ j \in \{ 1,2,\dots,N_{\rm{r}} \}}}
(|\mathbf{w}_j^H\mathbf{H}^H \mathbf{v}_i|^2).
\end{align}
Although exhaustive search with laborious beam sweeping can be used to find the optimal beam pair, it suffers from unacceptable overhead since there are a large number of beam pair candidates.

In this work, we aim to design a beam alignment method which collects detailed information about the channel in a hierarchical search manner before predicting the optimal beam. The DL techniques are employed to help design the PCs and predict the optimal beam. We shall show that the proposed method can be applied in both MISO and MIMO systems.



\section{Hierarchical Beam Alignment for the MISO System }
In this section, we consider the beam alignment problem for the MISO system. We propose an HBAN-MISO with two tiers of PCs to find the optimal beam in a coarse-to-fine search way.


\begin{figure*}[t]
\begin{centering}
\includegraphics[width=0.70 \textwidth]{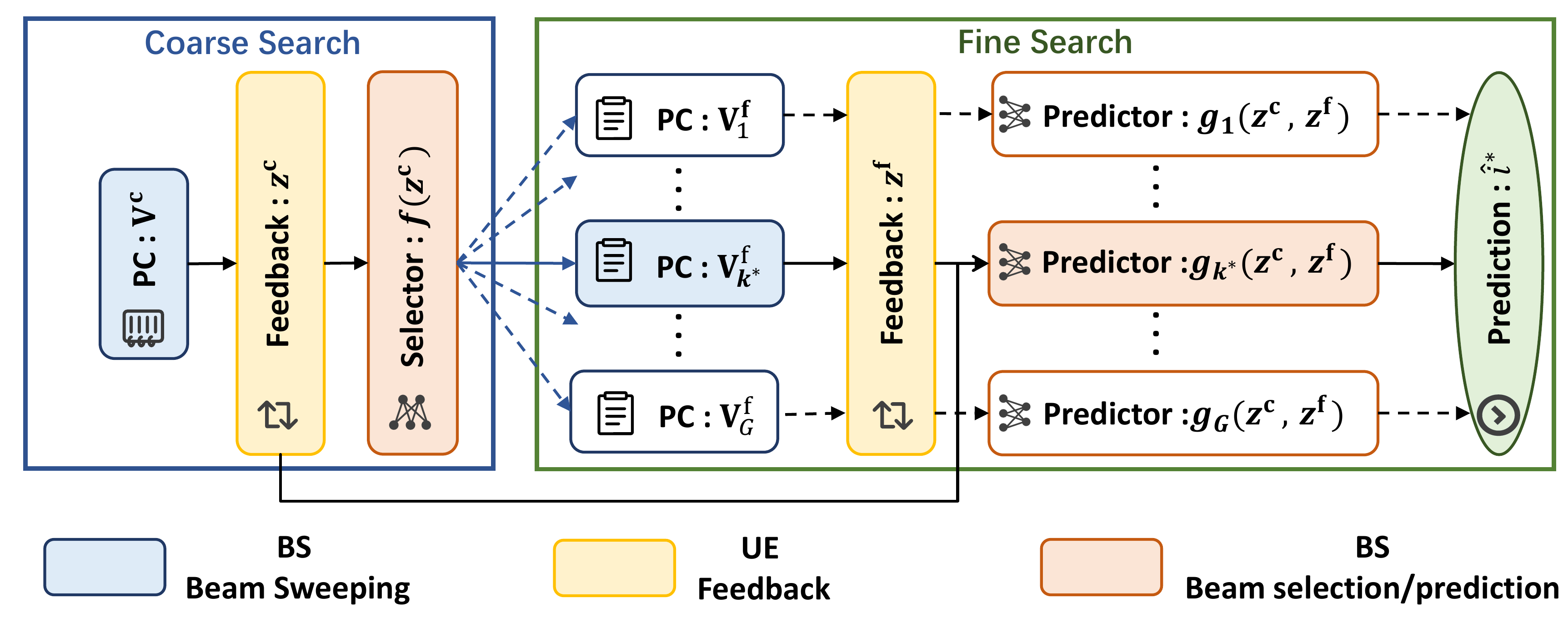}

\caption{The diagram of the proposed HBAN-MISO.}\label{architecture_single}
\end{centering}

\end{figure*}

\subsection{Hierarchical Beam Alignment Framework}
The proposed HBAN for the MISO system falls into the beam sweeping framework, where the BS sweeps PCs and then finds the optimal beam in the DFT codebook $\mathbf{V}$ with the probed channel information. To obtain a comprehensive measurement of the channel, the PCs have a hierarchical structure, where the first tier contains one coarse-search codebook and the second tier contains $G$ fine-search codebooks. Let $\mathbf{V}^{\rm{c}} = [\mathbf{v}^{\rm{c}}_{1},\mathbf{v}^{\rm{c}}_{2}, \dots, \mathbf{v}^{\rm{c}}_{N_1}] \in \mathbb{C}^{M_{\rm{t}} \times N_1}$ denote the coarse-search codebook with $\mathbf{v}^{\rm{c}}_{i}$ being the $i$th beam in the codebook and $N_1$ representing the codebook size. Likewise, let $\{\mathbf{V}^{\rm{f}}_1,\mathbf{V}^{\rm{f}}_2,\dots,\mathbf{V}^{\rm{f}}_G\}$ denote the set of fine-search codebooks with each $\mathbf{V}^{\rm{f}}_k = [\mathbf{v}^{\rm{f}}_{k,1},\mathbf{v}^{\rm{f}}_{k,2},\dots,\mathbf{v}^{\rm{f}}_{k,N_2}] \in \mathbb{C}^{M_{\rm{t}} \times N_2}$, where $k \in \{1,2,\dots,G\}$ is the index of the fine-search codebook. Here, we assume the $G$ fine-search codebooks have the same size $N_2$. The sum of PC sizes, $N_1$ + $N_2$, is usually much smaller than the DFT codebook size $N_{\rm{t}}$. With the two tiers of PCs, the proposed method performs beam alignment in a coarse-to-fine manner.


In the first step, the BS sweeps the coarse-search codebook $\mathbf{V}^{\rm{c}}$. The UE measures the received power, then reports the measurement to the BS, which is given as
\begin{equation}\label{equ:feedback}
    \mathbf{z}^{\rm{c}} = \left[|y^{\rm{c}}_{1}|^2,|y^{\rm{c}}_{2}|^2, \dots, |y^{\rm{c}}_{N_1}|^2\right]^T,
\end{equation}
where $y^{\rm{c}}_{i} = \sqrt{\rho}\mathbf{h}^H\mathbf{v}^{\rm{c}}_{i} s + n^{\rm{c}}$ is the received signal of beam $\mathbf{v}^{\rm{c}}_{i}$ at the UE. Based on the reported $\mathbf{z^{\rm{c}}}$, the BS utilizes a selector $f(\cdot)$ to select one of the $G$ fine-search codebooks, denoted as $\mathbf{V}^{\rm{f}}_{k^*}$, where $k^*$ is the index of the selected fine-search codebook. 

In the second step, the BS sweeps the selected fine-search codebook $\mathbf{V}^{\rm{f}}_{k^*}$. The UE also reports the corresponding power of the received signals to the BS, which is denoted as $\mathbf{z}^{\rm{f}} = \left[|y^{\rm{f}}_{k^*,1}|^2,|y^{\rm{f}}_{k^*,2}|^2, \dots, |y^{\rm{f}}_{k^*,N_2}|^2\right]^T$ with $y^{\rm{f}}_{k^*,i} = \sqrt{\rho}\mathbf{h}^H\mathbf{v}^{\rm{f}}_{k^*,i} s + n^{\rm{f}}$. Finally, the optimal beam can be predicted by the associated beam predictor $g_{k^*}(\cdot,\diamond)$ of the fine-search codebook $\mathbf{V}_{k^*}^{\rm{f}}$ based on the measurements of both $\mathbf{z}^{\rm{c}}$ and $\mathbf{z}^{\rm{f}} $. Our task is to employ the DL techniques to jointly design the two tiers of PCs $\{\mathbf{V}^{\rm{c}},\{\mathbf{V}^{\rm{f}}_{k}\}_{k=1}^{G}\}$, the selector $f(\cdot)$, and the beam predictors $\{g_k(\cdot,\diamond)\}_{k=1}^{G}$ so as to minimize the beam prediction error.

Compared with the method proposed in \cite{Jeffrey}, our method adopts the hierarchical structure and hence can significantly enrich the PC space without increasing the measurement overhead for each UE.
For the multi-UE scenario, the BS may sweep more probing beams under our hierarchical framework. In the worst case, the BS has to sweep all the probing beams and the number is $N_1 + GN_2$. If $G$ is not large, the increase of the sweeping overhead is tolerable.

In what follows, the specific DNN architecture with learnable two tiers of PCs is introduced.

\subsection{Deep Neural Network Architecture}
\begin{figure*}[t]
\begin{centering}
\includegraphics[width=0.6 \textwidth]{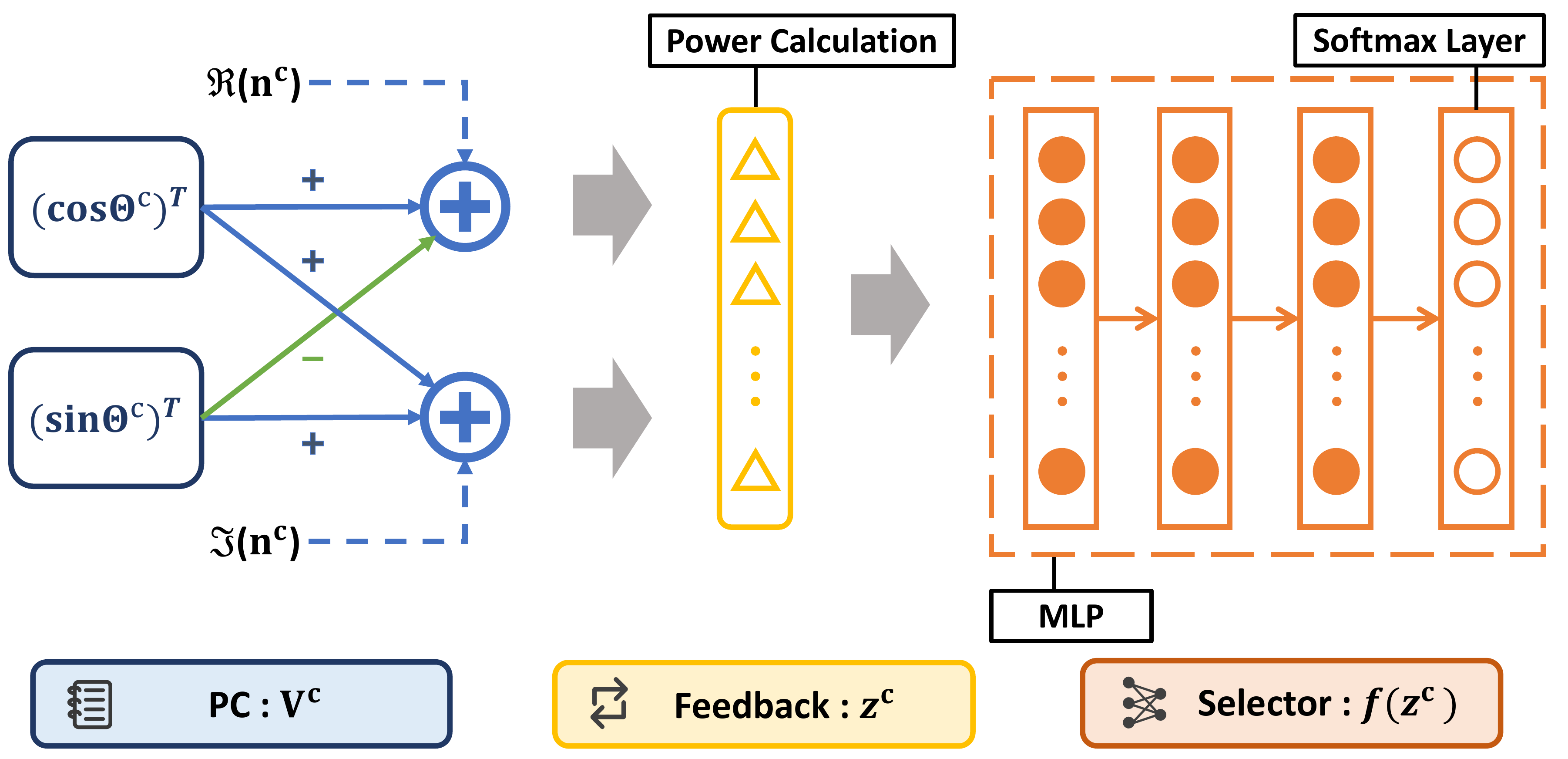}

\caption{The DNN architecture of the coarse-search part in HBAN-MISO.}\label{architecture_single_part}
\end{centering}

\end{figure*}

As shown in Fig. \ref{architecture_single}, the proposed HBAN-MISO is realized by the DNN including a coarse-search part and a fine-search part.
In the coarse-search part, the coarse-search codebook is modeled as a complex NN layer to calculate the received signal $\mathbf{y}^{\rm{c}} = \sqrt{\rho}(\mathbf{V}^{\rm{c}})^T\mathbf{h}^{*} + \mathbf{n}^{\rm{c}} \in \mathbb{C}^{N_1 \times 1}$.
Due to the constant-modulus constraint on each element in $\mathbf{V}^{\rm{c}}$, i.e., $|v^{\rm{c}}_{i,j}| = \frac{1}{\sqrt{M_{\rm{t}}}}$, the PC can be rewritten as
\begin{align}
    \mathbf{V}^{\rm{c}} = \frac{1}{\sqrt{M_{\rm{t}}}}\left[\cos(\pmb{\Theta}^{\rm{c}}) + \rm{j}\cdot\sin(\pmb{\Theta}^{\rm{c}})\right],
\end{align}
where $\pmb{\Theta}^{\rm{c}} \in \mathbb{R}^{M_{\rm{t}} \times N_1}$ is in fact the trainable parameter in the complex NN layer. As such, the calculation in the PC layer can be expressed as
\begin{align}\label{equ:single_z=wh}
    \begin{bmatrix}
             \Re\{\mathbf{y}^{\rm{c}}\} \\
             \Im\{\mathbf{y}^{\rm{c}}\}
    \end{bmatrix}
    = \sqrt{\rho}&\begin{bmatrix}
             (\cos(\pmb{\Theta}^{\rm{c}}))^T&-(\sin(\pmb{\Theta}^{\rm{c}}))^T \\
             (\sin(\pmb{\Theta}^{\rm{c}}))^T&~~(\cos(\pmb{\Theta}^{\rm{c}}))^T
    \end{bmatrix}
    \begin{bmatrix}
             \Re\{\mathbf{h}^{*}\} \\
             \Im\{\mathbf{h}^{*}\}
    \end{bmatrix} \notag \\
    + &\begin{bmatrix}
             \Re\{\mathbf{n}^{\rm{c}}\} \\
             \Im\{\mathbf{n}^{\rm{c}}\}
    \end{bmatrix},
\end{align}
where the noise $\mathbf{n}^{\rm{c}}$ satisfies $\mathcal{CN}(0,\sigma^2_n\mathbf{I})$. Then the measurement $\mathbf{z}^{\rm{c}}$ is calculated by following (\ref{equ:feedback}) in the feedback layer. 
The selector $f(\cdot)$ is realized by a multilayer perceptron (MLP) with one hidden layer, which outputs a likelihood vector $\mathbf{p} =\left[p_1,p_2,\dots,p_G\right]^T = f(\mathbf{z}^{\rm{c}}) \in \mathbb{R}^{G \times 1} $ to indicate the likelihood of fine-search codebooks. 
In contrast to the conventional hierarchical beam alignment method, we do not limit $N_1$ being equal to $G$.
The structure of DNN for the codebook and the operational process in the coarse-search part are shown in Fig.~\ref{architecture_single_part}. The index of the selected fine-search codebook is given as
\begin{equation}\label{equ:k^*}
    k^* = \mathop{\arg\max}\limits_{k \in \{ 1,2,\dots,G \} } p_{k}.
\end{equation}

In the fine-search part, all the $G$ fine-search codebooks are also modeled as complex NN layers. Similarly, we have $\mathbf{V}^{\rm{f}}_k = \frac{1}{\sqrt{M_{\rm{t}}}}\left[\cos(\pmb{\Theta}^{\rm{f}}_k) + \mathrm{j} \cdot \sin(\pmb{\Theta}^{\rm{f}}_k)\right]$ and $\pmb{\Theta}^{\rm{f}}_k \in \mathbb{R}^{M_{\rm{t}} \times N_2}$ is the trainable parameter in the $k$th fine-search codebook. With the selection result in (\ref{equ:k^*}), the measurement $\mathbf{z}^{\rm{f}}$ is derived through the $k^*$th fine-search PC layer and the feedback layer. The measurements $\mathbf{z}^{\rm{c}}$ and $\mathbf{z}^{\rm{f}}$ are both input to the associated beam predictor $g_{k^*}(\cdot,\diamond)$ to provide more detailed channel information and thus improve the prediction accuracy.
Each beam predictor is modeled as an MLP as well, which contains two hidden layers since the task of predicting the optimal beam is much more complex. In contrast to conventional two-tier hierarchical search methods that usually find the optimal beam from a subset of the DFT codebook $\mathbf{V}$ based on the previous searching results, the proposed beam predictor gives a likelihood $\mathbf{q} =\left[q_1,q_2,\dots,q_{N_{\rm{t}}}\right]^T = g_{k^*}(\mathbf{z}^{\rm{c}},\mathbf{z}^{\rm{f}}) \in \mathbb{R}^{N_{\rm{t}} \times 1}$ of all the beams in the DFT codebook $\mathbf{V}$ based on $\mathbf{z}^{\rm{c}}$ and $\mathbf{z}^{\rm{f}}$. 
In this way, the performance loss caused by decision error in the coarse-search part may be compensated.
Finally, the index of the optimal beam is decided as
\begin{equation}\label{equ:i_V_hat}
    \hat{i}^{*} = \mathop{\arg\max}\limits_{i \in \{ 1,2,\dots,N_{\rm{t}} \} } q_i.
\end{equation}

\subsection{Network Training}
Note that the proposed DNN is site-specific and needs to be retrained if the channel statistics change. In practice, the channel environment usually evolves slowly and remains almost static in a long period, indicating that there is no need to execute the retraining operation frequently. 

Here, a dataset $\mathcal{H}$ which contains a large amount of channel vectors $\mathbf{h}$'s is adopted in the training phase.
By simulations, we find that the end-to-end training strategy usually makes the DNN converge to a bad local optimal point. Thus we propose to first train the $\mathbf{V}^{\rm{c}}$ and $f(\cdot)$, then $\{\mathbf{V}^{\rm{f}}_{k}\}_{k=1}^{G}$ and $\{g_k(\cdot,\diamond)\}_{k=1}^{G}$ are trained based on the learned $\mathbf{V}^{\rm{c}}$ and $f(\cdot)$. Under this two-step training strategy, the proposed DNN can achieve a better performance.



\subsubsection{Training of the coarse-search codebook and the selector}
In the first training step, we train the coarse-search codebook $\mathbf{V}^{\rm{c}}$ and the selector $f(\cdot)$ jointly to make a coarse estimation of the channel.
The input is each channel sample $\mathbf{h} \in \mathcal{H}$ and the output is the likelihood vector $\mathbf{p}=\left[p_1,p_2,\dots,p_G\right]^T \in \mathbb{R}^{G \times 1}$ for the $G$ fine-search codebooks to be used. 

To facilitate the loss function design, we need to obtain the ground-truth label for each channel sample $\mathbf{h}$. Recall that if the BS employs the beams with close directions to the optimal DFT beam for a UE, such misalignment will not result in large SNR degradation. With the limited measurements, when the misalignment occurs, the predicted beam is usually close to the optimal beam. Thus a fine search concentrating on a smaller beam space can improve the performance.
Inspired by this, we propose to perform channel grouping by clustering channel samples with close optimal DFT beam directions.
Specifically, we first employ an oversampled DFT codebook $\mathbf{U}$ whose size is much larger than $N_{\rm{t}}$. Then the optimal DFT beam direction of each channel sample $\mathbf{h}$ in codebook $\mathbf{U}$ is found by performing noise-free exhaustive search. We denote the optimal beam as $\mathbf{u^*} = [1,e^{\mathrm{j}\frac{2\pi d}{\lambda}\sin(\alpha)},\dots,e^{\mathrm{j}\frac{2\pi d}{\lambda}(M_{\rm{t}}-1)\sin(\alpha)}]^T$ , where $\alpha$ is the optimal discrete beam direction. Note that a larger codebook size $M_{\mathbf{U}}$ can help us find a more precise beam direction of the channel sample.
Then we use the K-means clustering method \cite{Kmeans} to divide all the channel samples into $G$ groups based on their optimal DFT beam directions, where the distance between two channel vectors is defined as $|\sin(\alpha_i) - \sin(\alpha_j)|$.
After that, the group index of each channel sample is mapped to a one-hot vector $\mathbf{p}^{\rm{h}} = [p^{\rm{h}}_1,p^{\rm{h}}_2,\dots,p^{\rm{h}}_G]^T \in \mathbb{R}^{G \times 1}$ as the ground-truth label for training, where only $p^{\rm{h}}_k=1$ and the other elements are all zero if the channel sample is in group $k$.

To evaluate the difference between the predicted probability distribution and the true probability distribution, the cross-entropy (CE) function is adopted as the loss function, which can be expressed as
\begin{equation}\label{equ:CE_f1}
    L(\mathbf{p}) = -\frac{1}{G}\sum_{k=1}^{G} p^{\rm{h}}_{k} \log \left( p_{k} \right).
\end{equation}

\subsubsection{Training of fine-search codebooks and beam predictors}
After the coarse-search part is well trained, the second training step is to train the fine-search codebooks $\{\mathbf{V}^{\rm{f}}_{k}\}_{k=1}^{G}$ and beam predictors $\{g_k(\cdot,\diamond)\}_{k=1}^{G}$ jointly. The input are the channel sample $\mathbf{h} \in \mathcal{H}$ and the associated output result $\mathbf{p}$ from the coarse-search part. The output is the likelihood vector $\mathbf{q}=\left[q_1,q_2,\dots,q_{N_{\rm{t}}}\right]^T \in \mathbb{R}^{N_{\rm{t}} \times 1}$ for $N_{\rm{t}}$ beams in the DFT codebook $\mathbf{V}$.
The ground-truth label in this part is the optimal beam index in the codebook $\mathbf{V}$ and can be generated by performing noise-free exhaustive search in $\mathbf{V}$. We also encode each label to a one-hot vector $\mathbf{q}^{\rm{h}} = [q^{\rm{h}}_1,q^{\rm{h}}_2,\dots,q^{\rm{h}}_{N_{\rm{t}}}]^T  \in \mathbb{R}^{N_{\rm{t}} \times 1}$ and utilize the CE function as the loss function.
Since the data can be easily obtained in the wireless communication systems, the proposed DNN can be well trained to fully exploit the channel environment.
The whole training procedure is outlined in Algorithm \ref{algorithm1}.

\section{Hierarchical Beam Alignment for the MIMO System}

This section considers the MIMO system where beam alignment is supposed to be performed at both the BS and the UE. 
In particular, we aim to find the optimal BS-UE beam pair under the proposed DL-based hierarchical beam alignment framework, where a DNN of HBAN-MIMO is developed for beam pair prediction.


\begin{algorithm}[t]
\begin{algorithmic}[1]
\caption{Training Procedure of HBAN-MISO}\label{algorithm1}

\State \textbf{Input:} $\mathbf{h}$, $\mathbf{p}^{\rm{h}}$ and $\mathbf{q}^{\rm{h}}$
\State \textbf{\{Training for $\{\mathbf{V}^{\rm{c}},f(\cdot)\}$\}}
\State Utilize the K-means method for channel clustering and then generate the indicator vector $\mathbf{p}^{\rm{h}}$ for each channel sample $\mathbf{h}$;
\State Utilize $\{(\mathbf{h},\mathbf{p}^{\rm{h}})\}$ as the (feature, label) pair to learn $\mathbf{V}^{\rm{c}}$ and $f(\cdot)$ with the CE loss;

\State \textbf{\{Training for $\{\mathbf{V}^{\rm{f}}_{k}, g_k(\cdot,\diamond)\}_{k=1}^{G}$\}}
\State Select the fine-search codebook $\mathbf{V}^{\rm{f}}_{k^*}$ and the beam predictor $g_{k^*}(\cdot,\diamond)$ based on the output of  $f(\mathbf{z}^{\rm{c}})$;
\State Utilize $\{(\mathbf{h},\mathbf{q}^{\rm{h}})\}$ to learn $\mathbf{V}^{\rm{f}}_{k^*}$ and $g_{k^*}(\cdot,\diamond)$ with the CE loss;

\end{algorithmic}
\end{algorithm}

\subsection{Hierarchical Beam Pair Alignment Framework}

In MIMO systems, the conventional beam alignment methods usually consist of two steps, where the optimal beams at the BS and the UE are found in turn. Specifically, these methods first set omnidirectional beam pattern at the UE and seek the optimal beam at the BS based on swept measurement, then fix the optimal beam at the BS as the transmitting beam and sweep the probing beam at the UE to find the optimal beam. However, such two-stage methods usually have double sweeping overhead compared with beam alignment methods in the MISO system. Moreover, the methods are sensitive to the noise since they usually set omnidirectional beam pattern at the UE when seeking the optimal beam at the BS, which lacks beamforming gain.
To save the sweeping overhead and enhance the performance, we propose to perform beam pair alignment under the hierarchical search framework.
As such, every probing codeword is designed as the beam pair at the BS and the UE. In the beam pair searching procedure, the BS and the UE sweep the probing beam in each codeword of PCs to collect the channel information and finally predict the optimal beam pair with all measurements. Let $\Omega^{\rm{c}} = \{\mathcal{X}^{\rm{c}}_{1},\mathcal{X}^{\rm{c}}_{2}, \dots, \mathcal{X}^{\rm{c}}_{N_1}\}$ denote the coarse-search codebook with $N_1$ codewords, each of which is defined as $\mathcal{X}^{\rm{c}}_{i} = \{\mathbf{v}^{\rm{c}}_i,\mathbf{w}^{\rm{c}}_i \}$. In each codeword $\mathcal{X}^{\rm{c}}_{i}$, $\mathbf{v}^{\rm{c}}_i \in \mathbb{C}^{M_{\rm{t}} \times 1}$ and $\mathbf{w}^{\rm{c}}_i\in \mathbb{C}^{M_{\rm{r}} \times 1}$ denote the probing beams for the BS and the UE, respectively. 
Likewise, the set of fine-search codebooks can be defined as $\{\Omega^{\rm{f}}_{1},\Omega^{\rm{f}}_{2},\dots,\Omega^{\rm{f}}_{G}\}$, where $\Omega^{\rm{f}}_{k} = \{\mathcal{X}^{\rm{f}}_{k,1},\mathcal{X}^{\rm{f}}_{k,2}, \dots, \mathcal{X}^{\rm{f}}_{k,N_2}\}$ with $k \in \{1,2,\dots,G\}$ being the index and $\mathcal{X}^{\rm{f}}_{k,i} = \{\mathbf{v}^{\rm{f}}_{k,i},\mathbf{w}^{\rm{f}}_{k,i}\}$. Here, we also assume that there are $G$ fine-search codebooks in the second tier and all of them have the same size $N_2$. 
With such a redesigned PC structure, the proposed HBAN-MIMO can perform beam pair alignment for the MIMO system in a coarse-to-fine manner.


The beam alignment can be performed with the hierarchical PCs by two steps. In the first step, the BS and the UE sweep the coarse-codebook and measure the received power at the UE. We remark that the BS and the UE have to form a particular beam pair when sweeping each codeword.
To be specific, the analog beamformer at the BS and the UE is respectively $\mathbf{v}^{\rm{c}}_1$ and $\mathbf{w}^{\rm{c}}_1$ when sweeping the first codeword, then changed to $\mathbf{v}^{\rm{c}}_2$ and $\mathbf{w}^{\rm{c}}_2$ to sweep the second codeword and so on. The expression of the measurement reported to the BS $\mathbf{z}^{\rm{c}} = \left[|y^{\rm{c}}_{1}|^2, \dots, |y^{\rm{c}}_{N_1}|^2\right]^T$ is the same as (\ref{equ:feedback}), where the received signal of $i$th codeword $\mathcal{X}^{\rm{c}}_i$ is calculated by $y^{\rm{c}}_{i} = \sqrt{\rho}(\mathbf{w}^{\rm{c}}_i)^H\mathbf{H}^H\mathbf{v}^{\rm{c}}_i s + (\mathbf{w}^{\rm{c}}_i)^H \mathbf{n}^{\rm{c}}$. Similar to the proposed method for MISO systems, there is also a selector $f(\cdot)$ and several predictors $\{g_k(\cdot,\diamond)\}_{k=1}^{G}$ introduced in the HBAN-MIMO. Here, the selector $f(\cdot)$ is utilized to select one of the fine-search codebooks based on the reported feedback $\mathbf{z}^{\rm{c}}$. We denote the selected fine-search codebook as $\Omega^{\rm{f}}_{k^*}$ with $k^*$ being the selected index. In the second step, the BS and the UE tend to collect more detailed channel information by sweeping the fine-search codebook $\Omega^{\rm{f}}_{k^*}$. Similar to the coarse-search part, each codeword in fine-search codebooks is also designed as a beam pair. The measurement of received power $\mathbf{z}^{\rm{f}} = \left[|y^{\rm{f}}_{k^*,1}|^2, \dots, |y^{\rm{f}}_{k^*,N_2}|^2\right]^T$ is then reported to the BS with $y^{\rm{f}}_{i} = \sqrt{\rho}({\mathbf{w}^{\rm{f}}_{k^*,i}})^H\mathbf{H}^H\mathbf{v}^{\rm{f}}_{k^*,i} s + ({\mathbf{w}^{\rm{f}}_{k^*,i}})^H \mathbf{n}^{\rm{f}}$. Finally, the associated beam predictor $g_{k^*}(\cdot,\diamond)$ is utilized to predict the optimal beam pair based on $\mathbf{z}^{\rm{c}}$ and $\mathbf{z}^{\rm{f}}$.

The main modification for the proposed HBAN-MIMO in this scenario is the redesign of the codeword structure so that the BS and the UE work cooperatively to sweep the PCs. In addition, each beam predictor $\{g_k(\cdot,\diamond)\}_{k=1}^{G}$ is proposed to output the optimal beam pair at the BS and the UE.

\begin{figure*}[t]
\begin{centering}
\includegraphics[width=0.70 \textwidth]{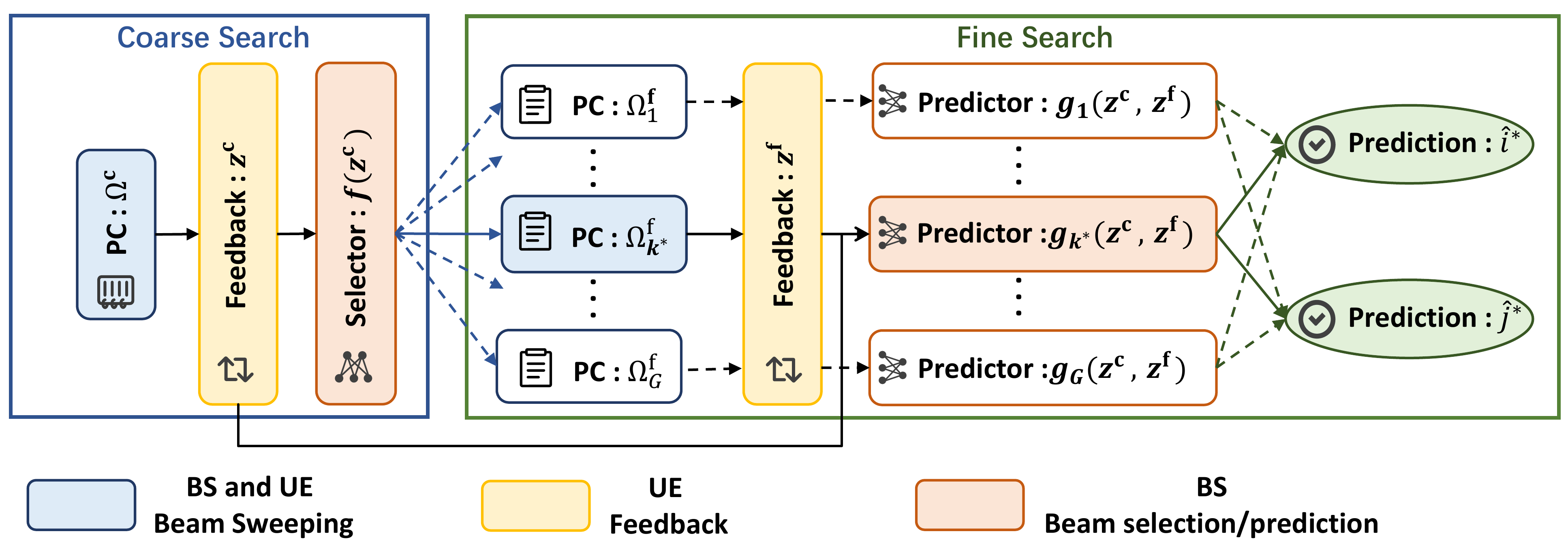}

\caption{The diagram of the proposed HBAN-MIMO.}\label{both_architecture}
\end{centering}

\end{figure*}

\subsection{Deep Neural Network Architecture}

\begin{figure*}[t]
\begin{centering}
\includegraphics[width=0.70 \textwidth]{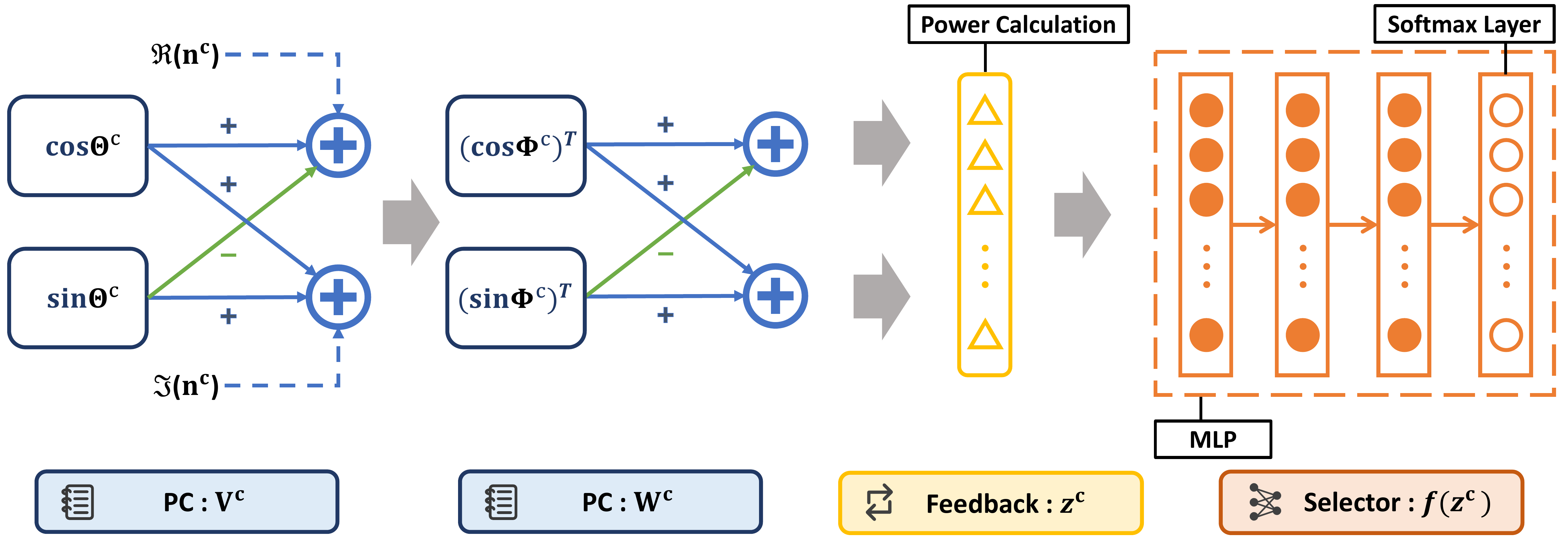}

\caption{The DNN architecture of the coarse-search part in HBAN-MIMO.}\label{architecture_both_part}
\end{centering}

\end{figure*}

The diagram of the proposed HBAN-MIMO is shown in Fig.  \ref{both_architecture}. The selector $f(\cdot)$ and predictors $\{g_k(\cdot,\diamond)\}_{k=1}^{G}$ are also modeled as MLPs and $f(\cdot)$ has the same structures as that in HBAN-MISO. 
Starting from the coarse-search part, we first express each codeword in the coarse-search codebook as
\begin{align} \label{equ:both_codeword}
\begin{split}
    \mathbf{v}^{\rm{c}}_i = \frac{1}{\sqrt{M_{\rm{t}}}}\left[\cos(\pmb{\theta}^{\rm{c}}_\mathit{i}) + \rm{j}\cdot\sin(\pmb{\theta}^{\rm{c}}_\mathit{i})\right] \\
    \mathbf{w}^{\rm{c}}_i = \frac{1}{\sqrt{M_{\rm{r}}}}\left[\cos(\pmb{\phi}^{\rm{c}}_\mathit{i}) + \rm{j}\cdot\sin(\pmb{\phi}^{\rm{c}}_\mathit{i})\right]
\end{split}
\end{align}
where $\pmb{\theta}^{\rm{c}}_i \in \mathbb{R}^{M_{\rm{t}} \times 1}$ and $\pmb{\phi}^{\rm{c}}_i \in \mathbb{R}^{M_{\rm{r}} \times 1}$ are trainable parameters.
The probing beam matrix for the BS and the UE are represented as $\mathbf{V}^{\rm{c}} = \left[\mathbf{v}^{\rm{c}}_1,\mathbf{v}^{\rm{c}}_2,\dots,\mathbf{v}^{\rm{c}}_{N_1}  \right] \in \mathbb{C}^{M_{\rm{t}} \times N_1}$ and $\mathbf{W}^{\rm{c}} = \left[\mathbf{w}^{\rm{c}}_1,\mathbf{w}^{\rm{c}}_2,\dots,\mathbf{w}^{\rm{c}}_{N_1}  \right] \in \mathbb{C}^{M_{\rm{r}} \times N_1}$, respectively. The trainable parameters of the probing beams are also given as $\pmb{\Theta}^{\rm{c}} = \left[\pmb{\theta}^{\rm{c}}_1,\pmb{\theta}^{\rm{c}}_2,\dots,\pmb{\theta}^{\rm{c}}_{N_1} \right] \in \mathbb{R}^{M_{\rm{t}} \times {N_1}}$ and $\pmb{\Phi}^{\rm{c}} = \left[\pmb{\phi}^{\rm{c}}_1,\pmb{\phi}^{\rm{c}}_2,\dots,\pmb{\phi}^{\rm{c}}_{N_1} \right] \in \mathbb{R}^{M_{\rm{r}} \times {N_1}}$. 
The received signal of the probing beams can be calculated by $\mathbf{y}^{\rm{c}} = \text{diag}(\sqrt{\rho} (\mathbf{W}^{\rm{c}})^H \mathbf{H}^H \mathbf{V}^{\rm{c}}) +(\mathbf{W}^{\rm{c}})^H \mathbf{n}^{\rm{c}}$. Then the received signal $\mathbf{y}^{\rm{c}}$ in the real number field can be expressed as 
\begin{align}\label{equ:both_z=wh_1}
    \begin{bmatrix}
             \Re\{\mathbf{y}^{\rm{c}}\} \\
             \Im\{\mathbf{y}^{\rm{c}}\}
    \end{bmatrix}
    =&
    \begin{bmatrix}
             \text{diag}( \Re\{\mathbf{A}^{\rm{c}}\} ) \\
             \text{diag}( \Im\{\mathbf{A}^{\rm{c}}\} )
    \end{bmatrix} \notag \\
    +&
    \begin{bmatrix}
             (\cos(\pmb{\Phi}^{\rm{c}}))^T & -(\sin(\pmb{\Phi}^{\rm{c}}))^T \\
             (\sin(\pmb{\Phi}^{\rm{c}}))^T & ~~(\cos(\pmb{\Phi}^{\rm{c}}))^T
    \end{bmatrix}
    \begin{bmatrix}
             \Re\{\mathbf{n}^{\rm{c}}\} \\
             \Im\{\mathbf{n}^{\rm{c}}\}
    \end{bmatrix},  
\end{align}
where the real and imaginary part of $\mathbf{A}^{\rm{c}}$ are given in (\ref{equ:both_z=wh_2}).
\begin{figure*}[!b]
\hrulefill
\begin{align}\label{equ:both_z=wh_2}
\begin{split}
    \begin{bmatrix}
             \Re\{\mathbf{A}^{\rm{c}}\} \\
             \Im\{\mathbf{A}^{\rm{c}}\}
    \end{bmatrix}
    = \sqrt{\rho}
    \begin{bmatrix}
             (\cos(\pmb{\Phi}^{\rm{c}}))^T & -(\sin(\pmb{\Phi}^{\rm{c}}))^T \\
             (\sin(\pmb{\Phi}^{\rm{c}}))^T & ~~(\cos(\pmb{\Phi}^{\rm{c}}))^T
    \end{bmatrix}
    \begin{bmatrix}
             \Re\{\mathbf{H}^{H}\} &-\Im\{\mathbf{H}^{H}\} \\
             \Im\{\mathbf{H}^{H}\} & ~~\Re\{\mathbf{H}^{H}\}
    \end{bmatrix} 
    \begin{bmatrix}
             \cos(\pmb{\Theta}^{\rm{c}})  \\
             \sin(\pmb{\Theta}^{\rm{c}}) 
    \end{bmatrix}.
\end{split}
\end{align}
\end{figure*}
Here, the noise $\mathbf{n}^{\rm{c}}$ satisfies $\mathcal{CN}(0,\sigma^2_n\mathbf{I})$. After getting the received signal $\mathbf{y}^{\rm{c}}$, the UE sends the measurement $\mathbf{z}^{\rm{c}}$ in (\ref{equ:feedback}) to the BS. Based on the feedback, the selector $f(\cdot)$ at the BS outputs a likelihood vector $\mathbf{p}$ and then obtains $k^*$ based on (\ref{equ:k^*}). 
Note that the PC layers in HBAN-MIMO contain the learnable PCs at both the BS and the UE in the MIMO system, whose detailed structure for the coarse-search part is shown in Fig.~\ref{architecture_both_part}.

In the fine-search part, we can also get the received signals $\mathbf{y}^{\rm{f}}$ with the selected fine-search codebook by (\ref{equ:both_z=wh_1}). 
Based on both measurements $\mathbf{z}^{\rm{c}}$ and $\mathbf{z}^{\rm{f}}$, the associated predictor $g_{k^*}(\cdot,\diamond)$ can provide the likelihoods of all possible beam pairs in pre-defined DFT codebooks $\mathbf{V}$ and $\mathbf{W}$. 
In contrast to designing only one network to output the likelihoods of all beam pairs, we utilize two sub-networks of $g_{k^*}(\cdot,\diamond)$ to separately obtain their optimal beams and each sub-network has the same structure as the beam predictor in HBAN-MISO.
Concretely, two vectors $\mathbf{q}_{\rm{t}}=\left[q_{\rm{t},1},q_{\rm{t},2},\dots,q_{\rm{t},N_{\rm{t}}}\right]^T \in \mathbb{R}^{N_{\rm{t}} \times 1}$ and $\mathbf{q}_{\rm{r}}=\left[q_{\rm{r},1},q_{\rm{r},2},\dots,q_{\rm{r},N_{\rm{r}}}\right]^T \in \mathbb{R}^{N_{\rm{r}} \times 1}$ are obtained with $\mathbf{q}_{\rm{t}} =  g_{k^*}^t(\mathbf{z}^{\rm{c}},\mathbf{z}^{\rm{f}})$ and $\mathbf{q}_{\rm{r}} =  g_{k^*}^r(\mathbf{z}^{\rm{c}},\mathbf{z}^{\rm{f}})$, which represents the likelihoods of the beams in $\mathbf{V}$ and $\mathbf{W}$, respectively.
Then the index of optimal beams at the BS and the UE can be given as    
\begin{align}
\begin{split}
    \hat{i}^{*} = \mathop{\arg\max}\limits_{i \in \{ 1,2,\dots,N_{\rm{t}} \} } q_{\rm{t},\mathit{i}}, \\
    \hat{j}^{*} = \mathop{\arg\max}\limits_{j \in \{ 1,2,\dots,N_{\rm{r}} \} } q_{\rm{r},\mathit{j}}.
\end{split}    
\end{align}

\begin{algorithm}[t]
\begin{algorithmic}[1]
\caption{Training Procedure of HBAN-MIMO}\label{algorithm2}
\State \textbf{Input:} $\mathbf{H}$, $\mathbf{p}^{\rm{h}}$, $\mathbf{q}_{\rm{t}}^{\rm{h}}$ and $\mathbf{q}_{\rm{r}}^{\rm{h}}$
\State \textbf{\{Training for $\{\Omega^{\rm{c}},f(\cdot)\}$\}}
\State Utilize the two-dimensional K-means method for channel clustering and then generate the indicator vector $\mathbf{p}^{\rm{h}}$ for each channel sample;
\State Utilize $\{(\mathbf{H},\mathbf{p}^{\rm{h}})\}$ as the (feature, label) pair to learn $\Omega^{\rm{c}}$ and $f(\cdot)$ with the CE loss;
\State \textbf{\{Training for $\{\Omega^{\rm{f}}_{k}, g_k(\cdot,\diamond)\}_{k=1}^{G}$\}}
\State Select the fine-search codebook $\Omega^{\rm{f}}_{k^*}$ and the beam predictor $g_{k^*}(\cdot,\diamond)$ based on the output of  $f(\mathbf{z}^{\rm{c}})$;
\State Utilize $\{(  \mathbf{H},\mathbf{q}_{\rm{t}}^{\rm{h}},,\mathbf{q}_{\rm{r}}^{\rm{h}}  )\}$ to learn $\mathbf{V}^{\rm{f}}_{k^*}$ and $g_{k^*}(\cdot,\diamond)$ with the CE loss;
\end{algorithmic}
\end{algorithm}
\vspace{-0.8cm}




\subsection{Network Training}
The dataset $\mathcal{H}$ adopted in the training phase for the proposed DNN contains a large amount of channel matrix $\mathbf{H}$'s to cover the channel information as completely as possible.
To prevent the DNN from converging to bad local optimal points, the network training for the revised DNN also follows the two-step strategy. The training procedure of HBAN-MIMO is similar to that of HBAN-MISO and is outlined in Algorithm \ref{algorithm2}. However, there are some differences for implementation which are explained below. 
\subsubsection{Generation of the ground-truth label in the first training step}
In the first training step for the coarse-search part, the ground-truth label should be generated firstly by clustering to facilitate the loss function design. 
Since the BS and the UE are required to perform beam alignment simultaneously in MIMO systems, we propose to perform channel grouping by clustering channel samples with close optimal DFT beam direction pairs.
Specifically, two DFT codebooks $\mathbf{U}$ and $\mathbf{T}$ are introduced for the BS and the UE, respectively, whose sizes are assumed to be much lager than $N_{\rm{t}}$ and $N_{\rm{r}}$. For each channel sample $\mathbf{H}$, the optimal DFT beam pair $\{\mathbf{u}^*,\mathbf{t}^*\}$ in codebooks $\mathbf{U}$ and $\mathbf{T}$ is obtained by performing noise-free exhaustive search. Here, the single-side beams of the optimal beam pair are denoted as $\mathbf{u}^* = [1,e^{\mathrm{j}\frac{2\pi d}{\lambda}\sin(\alpha)},\dots,e^{\mathrm{j}\frac{2\pi d}{\lambda}(M_{\rm{t}}-1)\sin(\alpha)}]^T$ and $\mathbf{t}^* = [1,e^{\mathrm{j}\frac{2\pi d}{\lambda}\sin(\beta)},\dots,e^{\mathrm{j}\frac{2\pi d}{\lambda}(M_{\rm{r}}-1)\sin(\beta)}]^T$, where $\alpha$ and $\beta$ represent the optimal discrete beam directions of the BS and the UE, respectively. 
Then channel clustering is conducted with $(sin(\alpha),sin(\beta))$ under the two-dimensional K-means scheme. We divide all the channel samples into $G$ groups, where the distance between two channel samples is defined as $\sqrt{(sin(\alpha_i)-sin(\alpha_j))^2+(sin(\beta_i)-sin(\beta_j))^2}$. Finally, the group index of each channel sample is mapped to a one-hot vector $\mathbf{p}^{\rm{h}} = [p^{\rm{h}}_1,\dots,p^{\rm{h}}_G]^T \in \mathbb{R}^{G \times 1}$ as the ground-truth label for training.

\begin{figure*}[t]
\begin{centering}
\includegraphics[width=0.65 \textwidth]{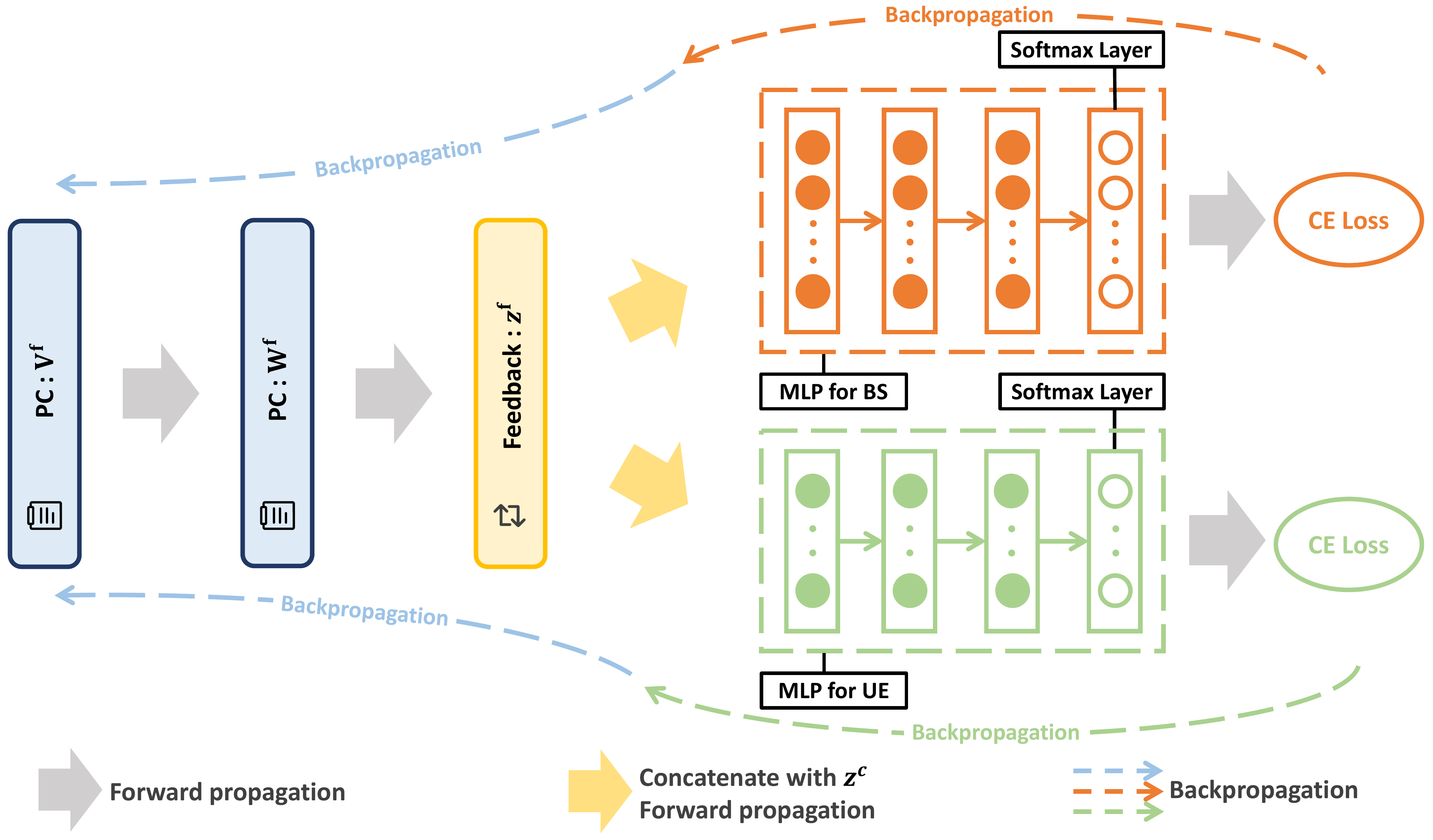}

\caption{The DNN architecture of the fine-search part in HBAN-MIMO.}\label{predictor_both}
\end{centering}

\end{figure*}
\subsubsection{Loss function for DNN training in the second training step}
The fine-search codebooks $\{\Omega^{\rm{f}}_{k}\}_{k=1}^{G}$ and beam predictors $\{g_k(\cdot,\diamond)\}_{k=1}^{G}$ are trained jointly in the second training step. In  the MIMO system, the ground-truth label is proposed to contain the optimal beam index in the pre-defined DFT codebook $\mathbf{V}$ as well as $\mathbf{W}$, which is generated by noise-free exhaustive search in advance.
The DNN structure of beam predictors are shown in Fig.~\ref{predictor_both}. 
After encoding the ground-truth label of each sample to the one-hot vectors $\mathbf{q}^{\rm{h}}_{\rm{t}}=\left[q^{\rm{h}}_{\rm{t},1},q^{\rm{h}}_{\rm{t},2},\dots,q^{\rm{h}}_{\rm{t},N_{\rm{t}}}\right]^T \in \mathbb{R}^{N_{\rm{t}} \times 1}$ and $\mathbf{q}^{\rm{h}}_{\rm{r}}=\left[q^{\rm{h}}_{\rm{r},1},q^{\rm{h}}_{\rm{r},2},\dots,q^{\rm{h}}_{\rm{r},N_{\rm{t}}}\right]^T \in \mathbb{R}^{N_{\rm{r}} \times 1}$, the loss function for the fine-search part can be expressed as
\begin{equation}\label{equ:CE_f2}
    L(\mathbf{p}) = -\frac{\xi}{N_{\rm{t}}}\sum_{k=1}^{N_{\rm{t}}} q^{\rm{h}}_{\rm{t},k} \log \left( q_{\rm{t},k} \right)-\frac{1-\xi}{N_{\rm{r}}}\sum_{k=1}^{N_{\rm{r}}} q^{\rm{h}}_{\rm{r},k} \log \left( q_{\rm{r},k} \right),
\end{equation}
where $\xi \in \left[0,1\right]$ is a tuning parameter in the training process which is derived by empirical results.

\begin{figure}[t]
\begin{centering}
\includegraphics[width=0.35 \textwidth]{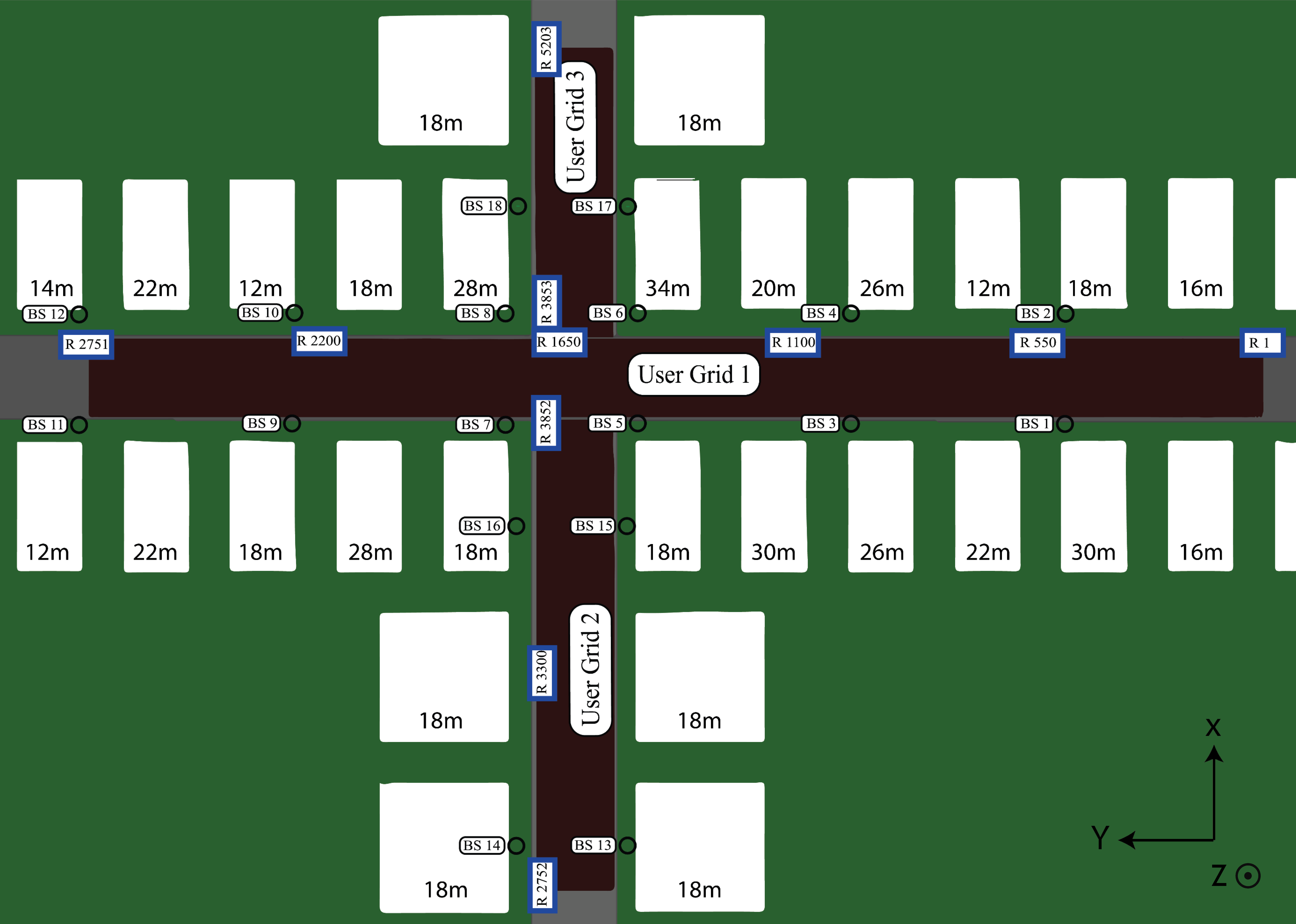}

\caption{The top view of DeepMIMO O1 scenario.}\label{DeepMIMO_O1}
\vspace{0.4cm}
\includegraphics[width=0.35 \textwidth]{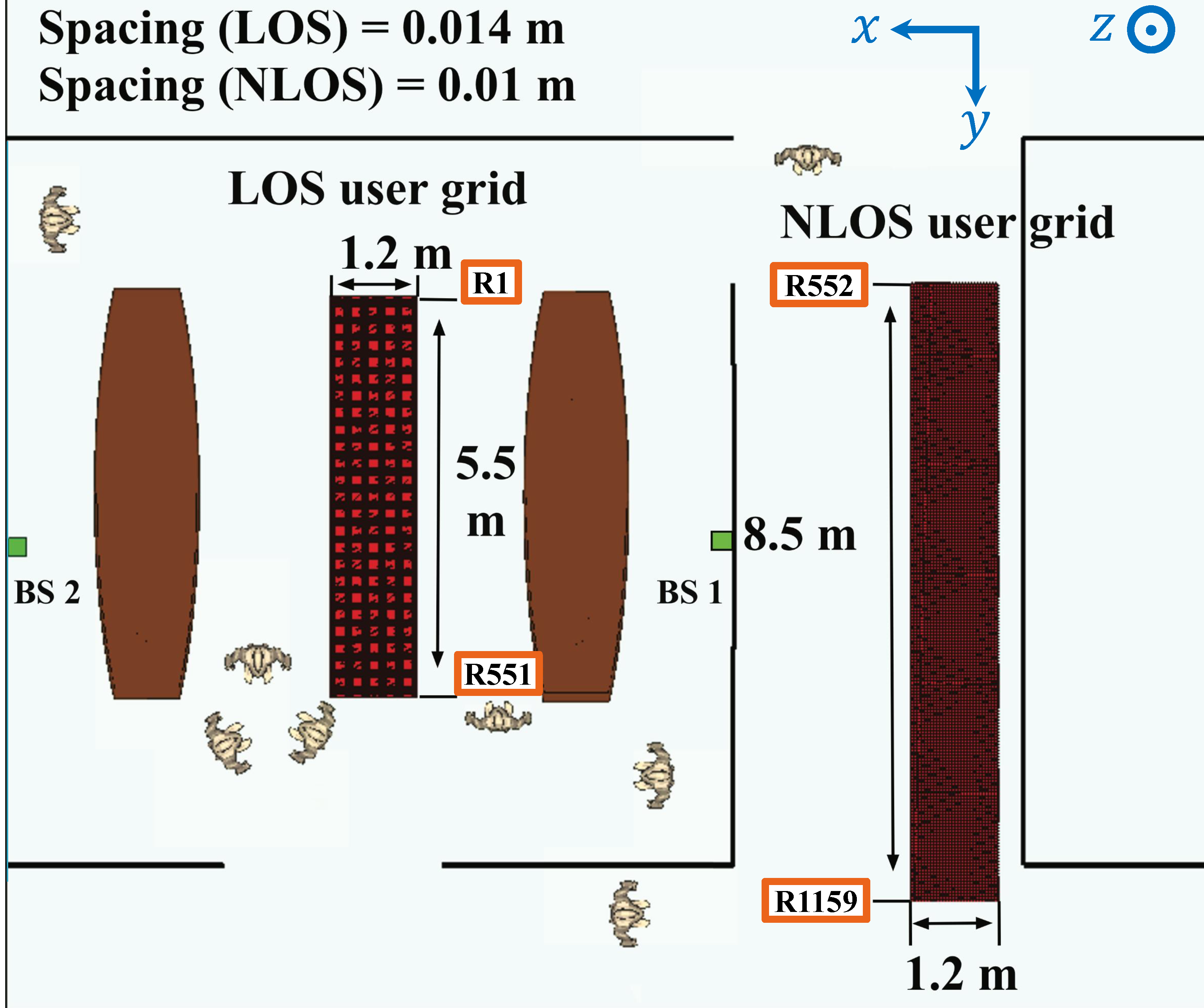}

\caption{The top view of DeepMIMO I3 scenario.}\label{DeepMIMO_I3}
\end{centering}
\end{figure}

\begin{figure}[t] 
\centering
\subfigure[DeepMIMO O1]
{\includegraphics[width=.40\textwidth]{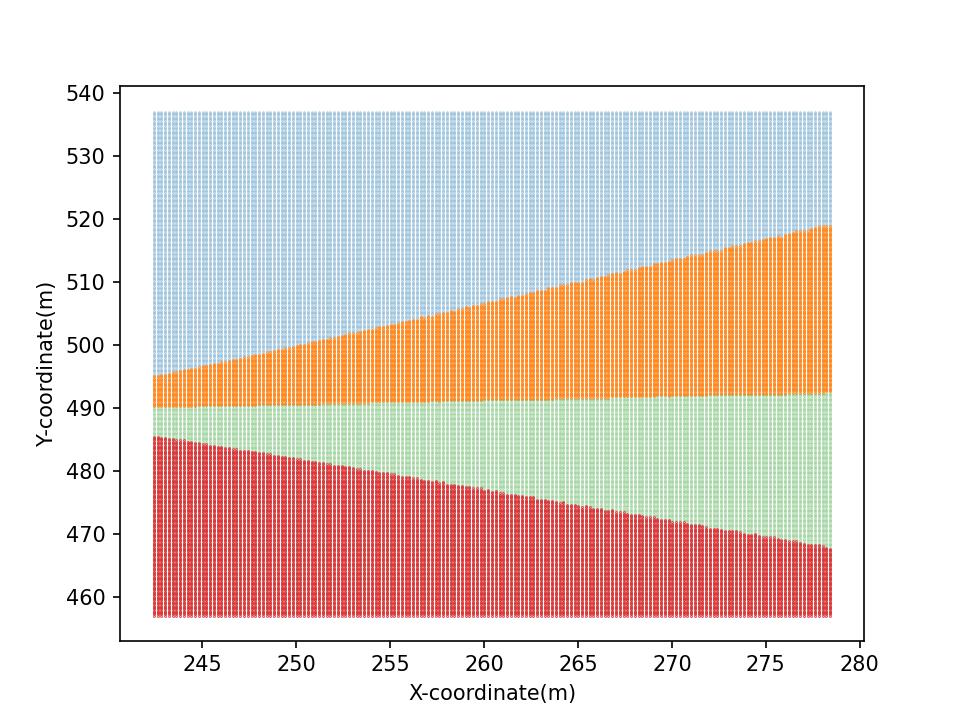}\label{clustering_single_a}}
\subfigure[DeepMIMO I3]
{\includegraphics[width=.40\textwidth]{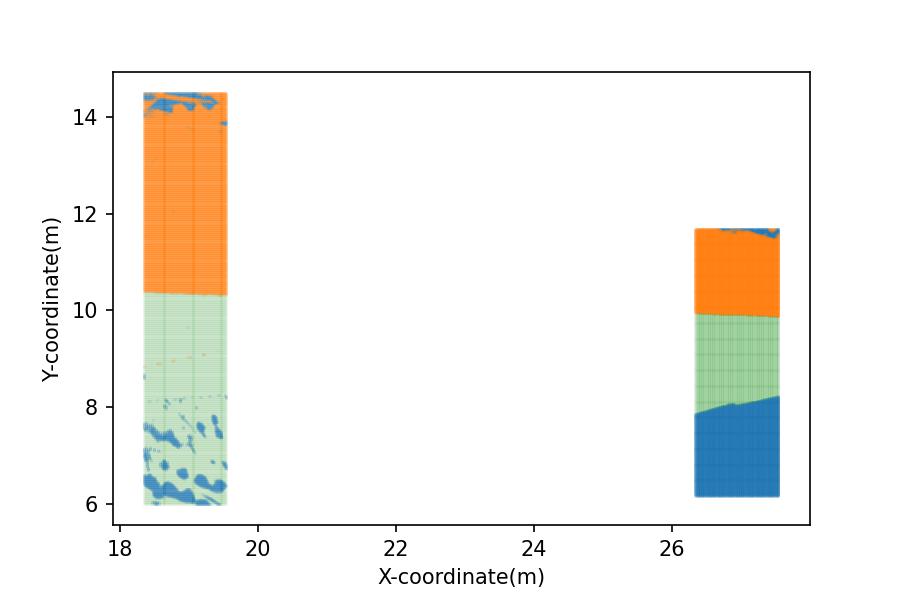}\label{clustering_single_b}}
\caption{Color-distinguished K-means clustering result of UE positions in MISO systems.}\label{clustering_single}
\end{figure}

\section{Simulation Results}
We perform extensive simulations based on the public datasets of DeepMIMO O1 and DeepMIMO I3 \cite{deepmimo} to compare the performance of our method and representative benchmarks. Both datasets are generated by ray-tracing.

\textbf{DeepMIMO O1 experiment:} The DeepMIMO O1 scenario shown in Fig.~\ref{DeepMIMO_O1}  simulates an outdoor urban environment with a few blocks and is available in the public DeepMIMO dataset \cite{deepmimo}. The BS \#3 and UEs in row \#800 to row \#1200 are selected for our simulation. There are a total of 72581 UEs evenly distributed in the street and most of them are  line-of-sight (LoS) UEs. The carrier frequency of the scenario is $28$ GHz.

\textbf{DeepMIMO I3 experiment:} The DeepMIMO I3 scenario shown in Fig.~\ref{DeepMIMO_I3} models an indoor conference room where several walls are placed. The BS\#2 which is placed on the wall inside the room is selected for our simulation. There are altogether 118959 UEs in this scenario which are placed inside the conference room or in the hallway. Here, the UEs in the conference room are mostly LoS UEs while those in the hallway are non-line-of-sight (NLoS) UEs. The carrier frequency of the scenario is $60$ GHz.

Here, for both scenarios, we use 60\% samples for training, 20\% for validation and 20\% for testing.

\subsection{MISO System}
We first consider the MISO system to evaluate the performance of our proposed hierarchical beam alignment method. The number of antenna elements at the BS is set to $M_{\rm{t}}=64$ and the antenna spacing is set to $d = \frac{\lambda}{2}$. The bandwidth and the noise power spectral density (PSD) are set to $100$ MHz and -161 dBm/Hz, respectively, if not specified otherwise. The transmit power is 10 dBm. The DFT codebook $\mathbf{V}$ for downlink transmission has $N_{\rm{t}}=128$ codewords.    
Based on the elbow method of the K-means clustering algorithm, we set the number of fine-search codebooks as $G=4$ in the DeepMIMO O1 experiment and $G=3$ for DeepMIMO I3. For the MLP that models $f(\cdot)$, parameters such as the input and output layer dimensions are determined based on $N_1$ and $G$, the size of the hidden layer is set to $N_1$ as well and thus the computational complexity is $O(N_1^2 + N_1G)$. For each MLP $g_k(\cdot)$, the input and output layer dimensions are $(N_1+N_2)$ and $N_{\rm{t}}$, we set the sizes of hidden layers to $2(N_1+N_2)$ and $3(N_1+N_2)$, thus the computational complexity is $O((N_1+N_2)^2 + (N_1+N_2)N_{\rm{t}})$.


\subsubsection{Channel clustering}
The clustering result for the two experiments are shown in Fig. \ref{clustering_single}, where the UE positions are marked as scattered points. All UEs are divided into different groups distinguished by colors. With the K-means clustering performed in the sine space, it is observed that the UEs in close proximity are grouped together and thus the fine search for each group will concentrate on a much smaller beam space. In the DeepMIMO I3 experiment, since a large amount of UEs only have NLoS paths, the UEs in different groups are not completely radially distributed.

\subsubsection{Performance evaluation}
We utilize the accuracy and spectral efficiency as performance metrics, where the former represents the probability of correctly predicting the optimal beam.
To verify the superiority of our method, the benchmarks for performance comparison are listed as follows.

\textbf{Method proposed in \cite{Jeffrey}:} The method adopts a learnable one-tier PC to collect the channel information, then utilizes a DNN to predict the optimal beam in the adopted DFT codebook $\mathbf{V}$.

\textbf{AMCF codebook based search:} The method adopts the same DNN architecture as the proposed HBAN-MISO, but the learnable probing codebook is replaced by the fixed two-tier AMCF codebook with the same codebook size.


\textbf{Exhaustive search:} The BS sweeps all the beams in the adopted DFT codebook $\mathbf{V}$ and then selects the beam with the highest received power for downlink transmission. 

\textbf{Two-tier hierarchical search:} This method utilizes a two-tier hierarchical codebook, where the tier-1 codebook consists of several wide beams while the tier-2 codebook contains all the narrow beams in the adopted DFT codebook $\mathbf{V}$. Moreover, each narrow beam in the tier-2 codebook is a child beam of one of the wide beams in the tier-1 codebook. The BS first sweeps wide beams in the tier-1 codebook then sweeps child beams of the best wide beam in the tier-2 codebook, finally the child beam with the highest received power is selected. Here, we adopt the AMCF proposed in \cite{hierarchical3} to generate wide beams for the method.

\begin{table}[t]
\newcommand{\tabincell}[2]{\begin{tabular}{@{}#1@{}}#2\end{tabular}}
\caption{Probing Codebook Sizes in DeepMIMO O1}\label{HBAN-MISO}
\vspace{-0.4cm}
\begin{center}

\begin{tabular}{c|c c c c c c c c }
\hline
$N_1$ & 3 & 4 & 4 & 6 & 6 & 6 & 6 &  6\\
\hline
$N_2$ & 3 & 4 & 6 & 6 & 8 & 10 & 12 & 14\\
\hline
Sum & 6 & 8 & 10 & 12 & 14 & 16 & 18 & 20\\
\hline

\end{tabular}
\label{tab1}
\end{center}
\vspace{-0.4cm}
\end{table}

\begin{table}[t]
\newcommand{\tabincell}[2]{\begin{tabular}{@{}#1@{}}#2\end{tabular}}
\caption{Probing Codebook Sizes in DeepMIMO I3}\label{MISO_benchmark}
\vspace{-0.4cm}
\begin{center}

\begin{tabular}{c|c c c c c c c c }
\hline
$N_1$ & 3 & 3 & 3 & 4 & 5 & 6 & 6 &  6\\
\hline
$N_2$ & 3 & 5 & 7 & 8 & 9 & 10 & 12 & 14\\
\hline
Sum & 6 & 8 & 10 & 12 & 14 & 16 & 18 & 20\\
\hline
\end{tabular}
\label{tab2}
\end{center}
\vspace{-0.4cm}
\end{table}

\begin{table}[t]
\newcommand{\tabincell}[2]{\begin{tabular}{@{}#1@{}}#2\end{tabular}}
\caption{Beam Sweeping Complexity of Different Methods in MISO systems}\label{MISO_complexity}
\vspace{-0.4cm}
\begin{center}

\begin{tabular}{c | c}

\hline
\hline
\textbf{Method} & \textbf{Complexity ($N_{\rm{t}}=128$)}  \\
\hline
HBAN-MISO & $N_1+N_2$  \\

Method proposed in \cite{Jeffrey} & $N_1+N_2$ \\

AMCF codebook based search & $N_1+N_2$ \\

Exhaustive search &  $128$ \\

Two-tier hierarchical search & $11+12=23$ \\

Binary search & $2\times\log_{2}{128}=14$ \\

\hline
\hline
\end{tabular}
\end{center}

\end{table}

\begin{figure}[t] 
\centering
\subfigure[DeepMIMO O1]
{\includegraphics[width=.40\textwidth]{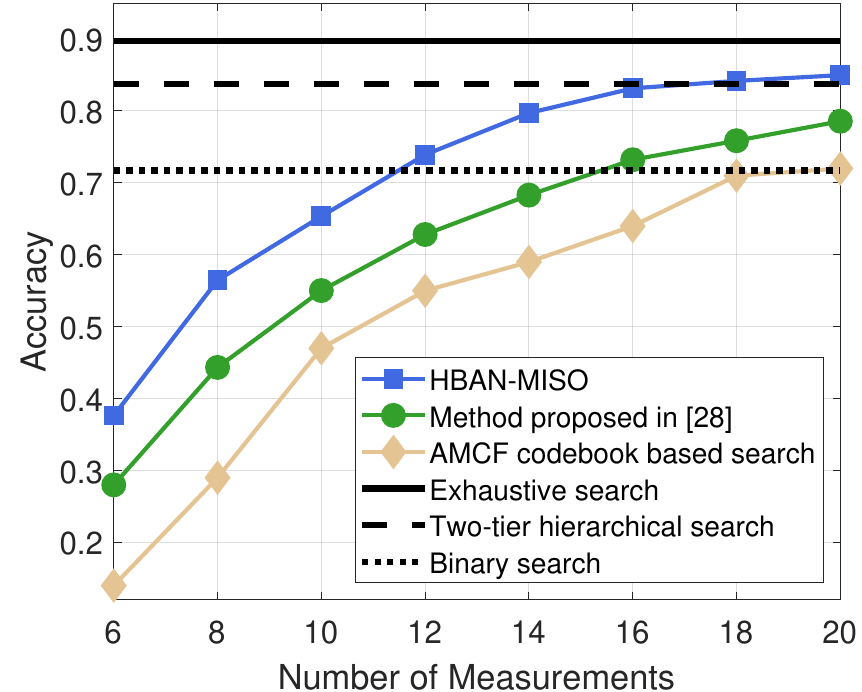}\label{deepmimo_O1_booksize_single}}
\subfigure[DeepMIMO I3]
{\includegraphics[width=.40\textwidth]{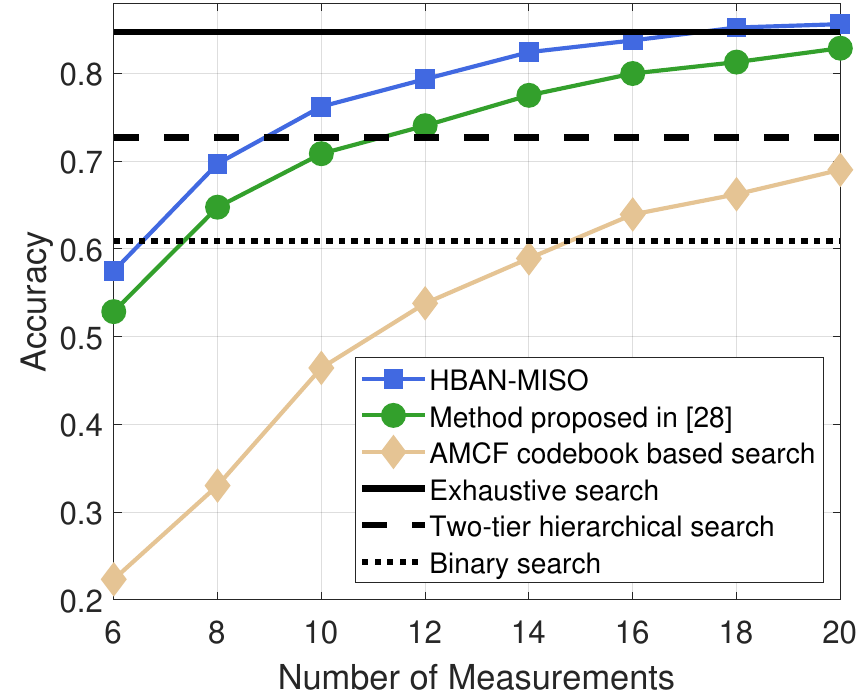}\label{deepmimo_I3_booksize_single}}
\caption{Beam alignment accuracy v.s. number of measurements in MISO systems.}\label{booksize_single}

\end{figure}

\textbf{Binary search:} The binary search can be seen as a special case of hierarchical search which performs a binary tree search on the adopted DFT codebook. For the DFT codebook $\mathbf{V}$ with $N_{\rm{t}}$ beams, the codebook in the method consists of $\log_2 N_{\rm{t}}$ tiers. In each tier, the BS splits the search space into two partitions of equal size by sweeping the corresponding wide beams and finally reaches sweeping the narrow beams in the DFT codebook $\mathbf{V}$. Similar to the two-tier hierarchical search, the AMCF is adopted to generate wide beams herein.



For fair comparison, all DL-based methods are evaluated with an equal number of measurements $(N_1+N_2)$, which directly determines the sweeping overhead of the proposed beam alignment method.
Under the condition with fixed probing overhead, the sizes of codebooks in the fine-search part should be no smaller than that in the coarse-search part, i.e., $N_1 \le N_2$. This is intuitive since a more precise prediction is required in the fine-search part and more detailed channel information is necessary. 
We set the sizes of codebooks in the two parts of the HBAN-MISO based on the simulation trials and show them in Table~\ref{HBAN-MISO} and Table~\ref{MISO_benchmark}.

Fig.~\ref{booksize_single} shows the average accuracy of HBAN-MISO and the benchmarks with respect to the number of measurements. As a remark, when $N_{\rm{t}}$ is fixed to 128, the exhaustive search and binary search have constant numbers of measurements, which are 128 and 14, respectively. 
For the two-tier hierarchical search, the BS needs to sweep a narrow-beam codebook in the second tier in addition to the wide-beam codebook in the first tier. Here, we denote the size of the wide-beam codebook as $N_{\rm{W}}$, then the size of each narrow-beam codebook can be calculated as $\lceil N_{\rm{t}}/N_{\rm{W}} \rceil$. It is noted that the number of measurements $N_{\rm{W}} + \lceil N_{\rm{t}}/N_{\rm{W}} \rceil$ fluctuates greatly as $N_{\rm{W}}$ changes, which makes it hard to compare the method with others fairly. Here, we show the performance of this benchmark with minimum sweeping complexity by setting $N_{\rm{W}}=\lfloor \sqrt{N_t} \rfloor = 11$ and $\lceil N_{\rm{t}}/N_{\rm{W}} \rceil=12$ in the experiment and observe whether HBAN-MISO can achieve comparable performance with less sweeping overhead. The beam sweeping complexity of all considered methods are shown in Table~\ref{MISO_complexity}.

\begin{table*}[t]
\renewcommand\arraystretch{1.5}
\newcommand{\tabincell}[2]{\begin{tabular}{@{}#1@{}}#2\end{tabular}}
\caption{Accuracy of HBAN-MISO and HBAN-MISO-PCS with different number of measurements in DeepMIMO O1}\label{step1_ACC}
\vspace{-0.4cm}
\begin{center}

\begin{tabular}{|l|c c c c c c c c|}

\hline
 
\diagbox[dir=NW]{Method}{ $N$\vspace{-0.4cm}} & 6 & 8 & 10 & 12 & 14 & 16 & 18 & 20 \\
\hline
HBAN-MISO-PCS & 0.427 & 0.591 & 0.743 & 0.768 & 0.831 & 0.861 & 0.880 & 0.885 \\
\hline
HBAN-MISO & 0.382 & 0.561 & 0.687 & 0.744 & 0.798 & 0.829 & 0.844 & 0.852 \\
\hline

\end{tabular}
\label{tab3}
\end{center}

\end{table*}
\begin{figure}[t] 
\centering
\subfigure[DeepMIMO O1]
{\includegraphics[width=.40\textwidth]{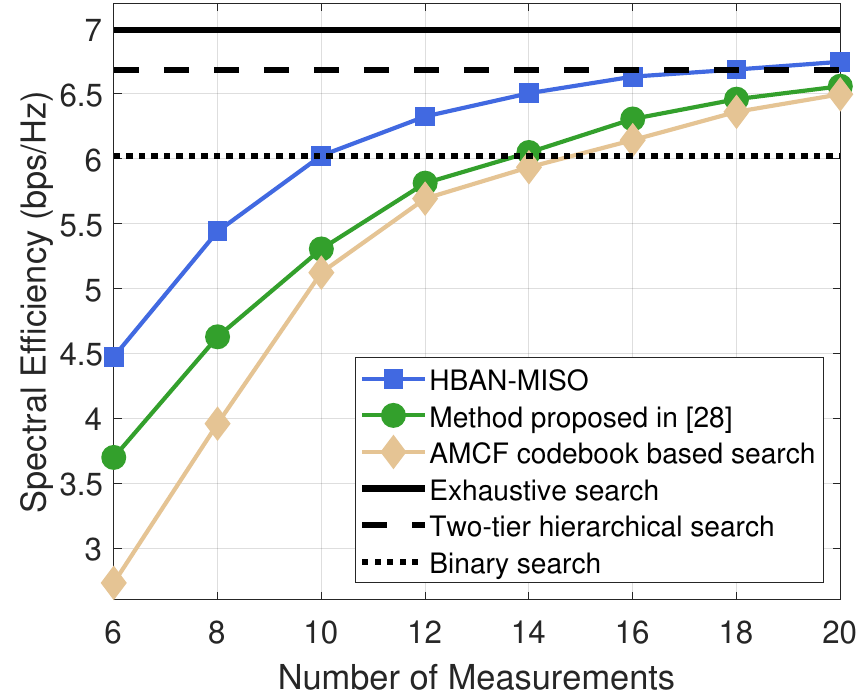}\label{deepmimo_O1_bps_single}}
\subfigure[DeepMIMO I3]
{\includegraphics[width=.40\textwidth]{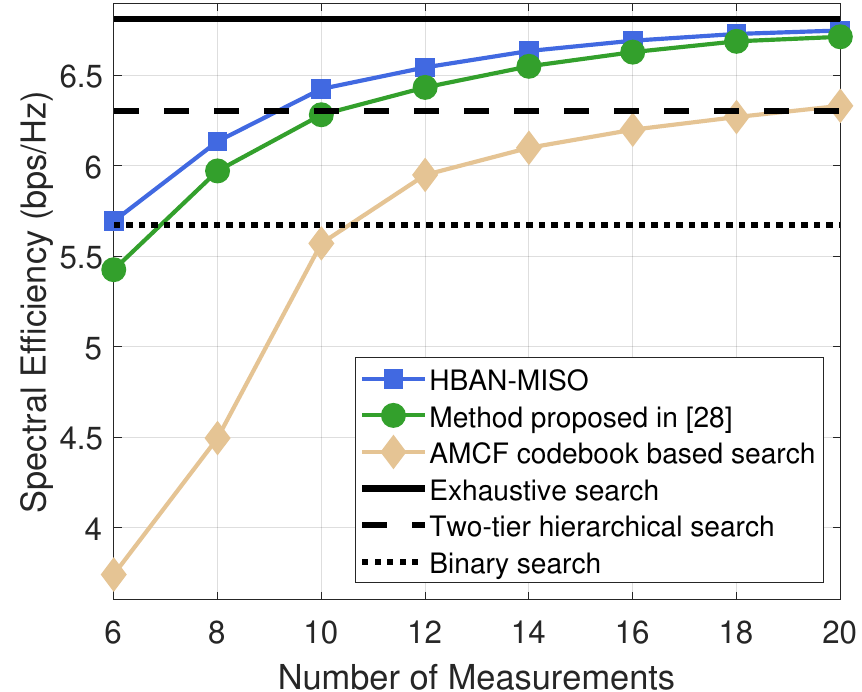}\label{deepmimo_I3_bps_single}}

\caption{Spectral efficiency v.s. number of measurements in MISO systems.}\label{bps_single}

\end{figure}

\begin{figure}[t] 
\centering
\subfigure[DeepMIMO O1]
{\includegraphics[width=.40\textwidth]{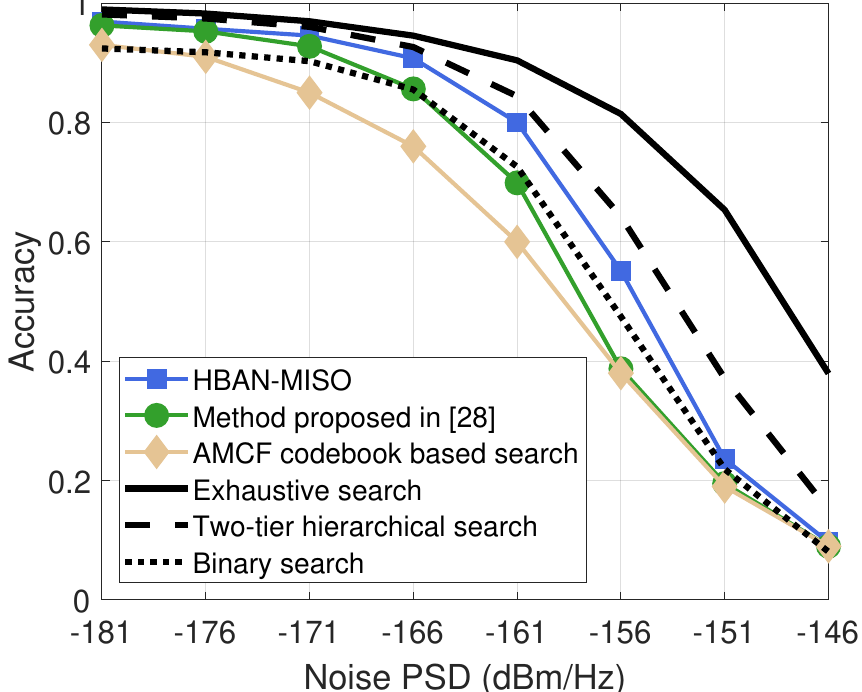}\label{fig5a}}
\subfigure[DeepMIMO I3]
{\includegraphics[width=.40\textwidth]{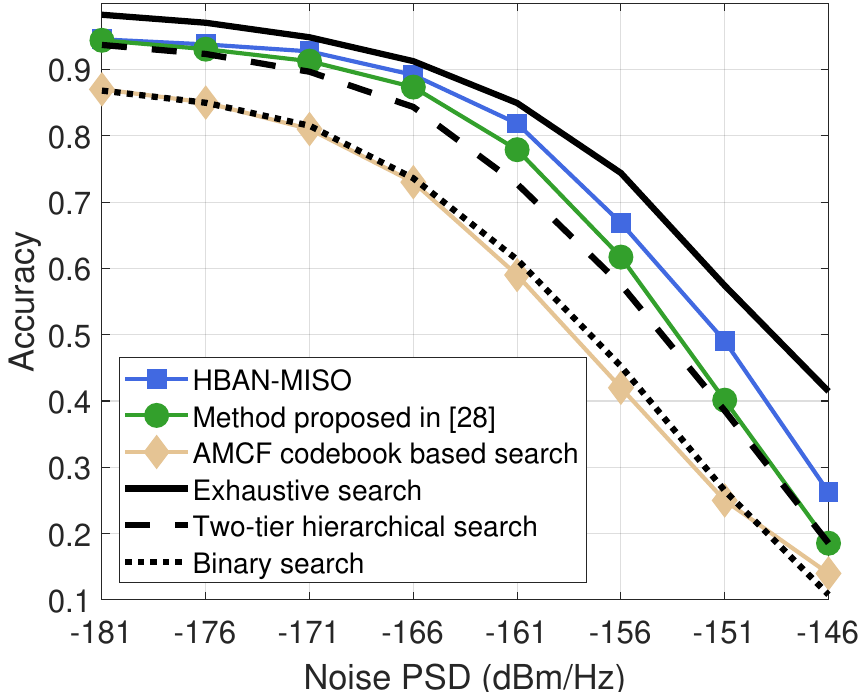}\label{fig5b}}

\caption{Beam alignment accuracy v.s. noise PSD in MISO systems.}\label{fig5}

\end{figure}

In Fig.~\ref{booksize_single}, we observe that when the number of measurements is small, the accuracy of HBAN-MISO is not impressive compared with conventional methods. However, as the measurement number increases, the performance of HBAN-MISO is greatly improved. This indicates that larger PCs can help the BS collect more detailed information about the channel. In addition, when the number of measurements is set to 20 in the DeepMIMO I3 scenario, the proposed method outperforms all benchmarks including the exhaustive search, this is due to the fact that the DL methods can also learn the statistical property of noise, which is lacking in conventional methods. 
It is also noteworthy that HBAN-MISO can consistently achieve a better performance than the method proposed in \cite{Jeffrey} and the AMCF codebook based search, thanks to the coarse-to-fine search mechanism and the learnable probing codebook, respectively.

We further analyze the beam alignment accuracy of the proposed HBAN-MISO and HBAN-MISO with perfect coarse search (HBAN-MISO-PCS), where the latter contains a perfect coarse-search part that can always obtain the accurate label from $\{1,2,\dots,G\}$ and serves as a performance upper bound. The simulation result in DeepMIMO O1 is shown in Table~\ref{step1_ACC}, from which it is observed that HBAN-MISO-PC only has a slight increase on the accuracy compared to HBAN-MISO, which indicates that the beam alignment inaccuracy of the proposed HBAN-MISO mainly comes from the fine-search part. Such conclusion is intuitive due to error accumulation and the fact that $G \ll N_{\rm{t}}$.

In practice, spectral efficiency is a main concern of the mmWave communication systems. Thus we also evaluate the impact of the measurements on the average spectral efficiency of our method and the benchmarks in Fig.~\ref{bps_single}. Similar to the accuracy performance, the DL-based methods can achieve substantially higher spectral efficiency with larger PCs. In addition, the proposed HBAN-MISO can achieve a performance close to the two-tier hierarchical search with only 18 and 10 measurements in DeepMIMO O1 and DeepMIMO I3, respectively.

To examine the robustness of considered methods, we evaluate the accuracy of these methods with different noise PSD in Fig.~\ref{fig5}. In the simulation, the number of measurements is set to $N_1 + N_2 = 14$ for fair comparison with the binary search method. The performance of all methods are gradually deteriorated as the noise PSD increases. When the noise PSD is smaller than -166 dBm/Hz, the performance of HBAN-MISO is similar to or even better than the two-tier hierarchical search with 39.2\% less sweeping overhead. Moreover, it can be observed that the HBAN-MISO always outperforms the method proposed in \cite{Jeffrey} and the AMCF codebook based search in different noise PSDs. Overall, we find that the learnable probing codebook based methods, i.e. HABN-MISO and the method proposed in \cite{Jeffrey}, can achieve a much better performance than conventional hierarchical search methods in the DeepMIMO I3 experiment, which indicates the robustness of the former.

\begin{figure*}[t]
  \centering
  \subfigure[$\mathbf{V}^{\rm{c}}$]
  {\includegraphics[width=.16\textwidth]{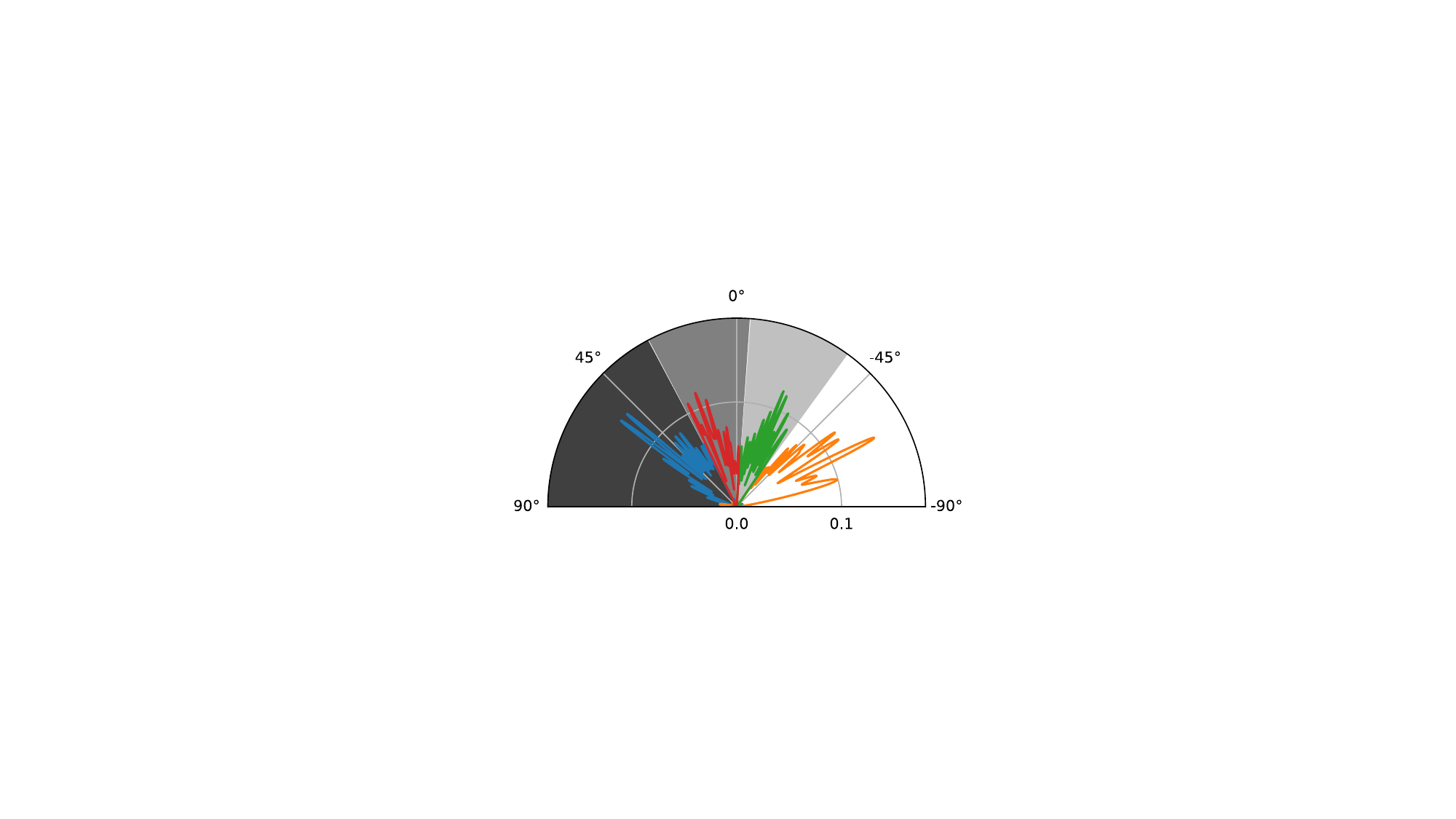}\label{fig6a}}
  \subfigure[$\mathbf{V}^{\rm{f}}_{1}$]
  {\includegraphics[width=.18\textwidth]{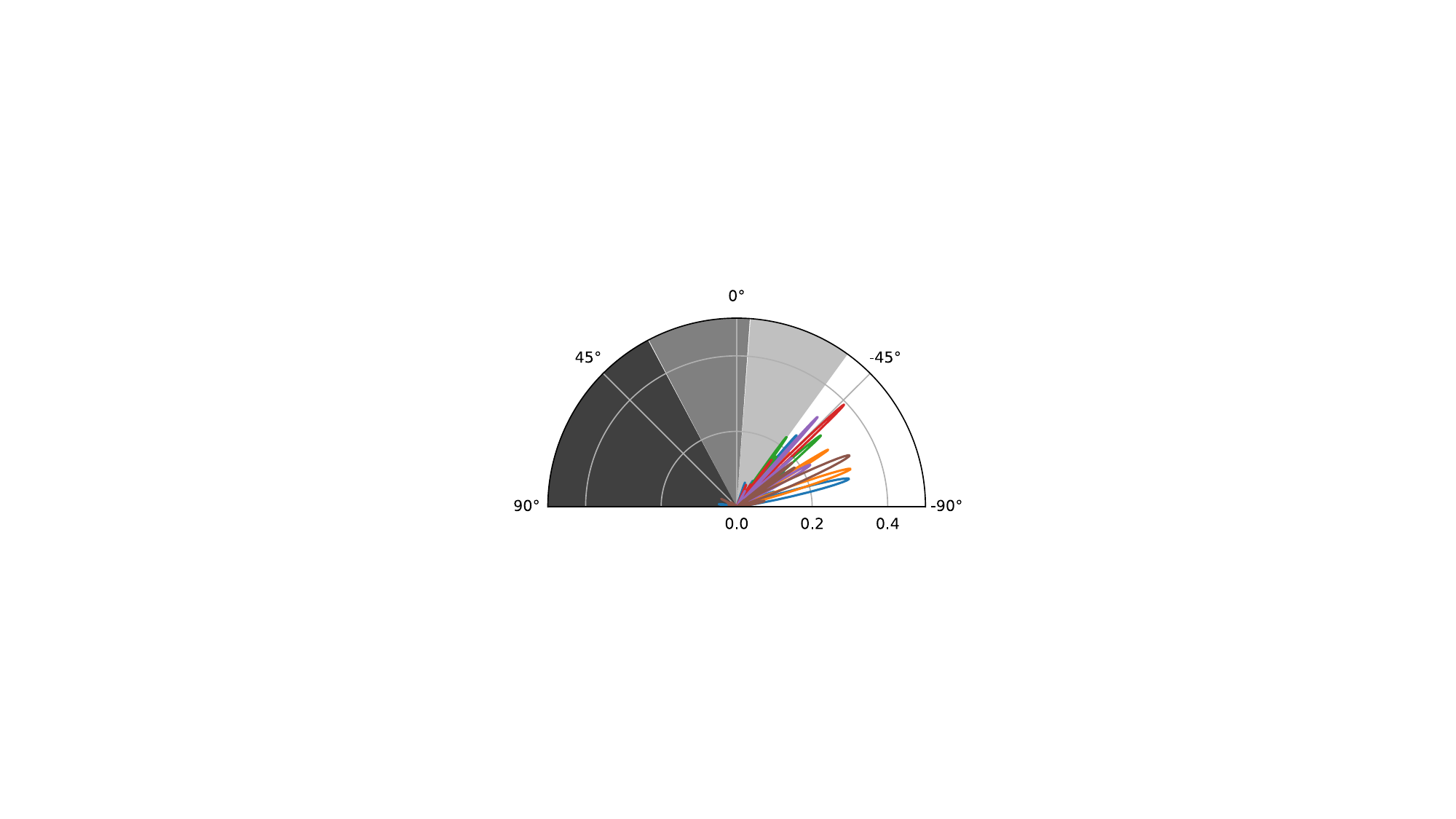}\label{fig6b}}
  \subfigure[$\mathbf{V}^{\rm{f}}_{2}$]
  {\includegraphics[width=.18\textwidth]{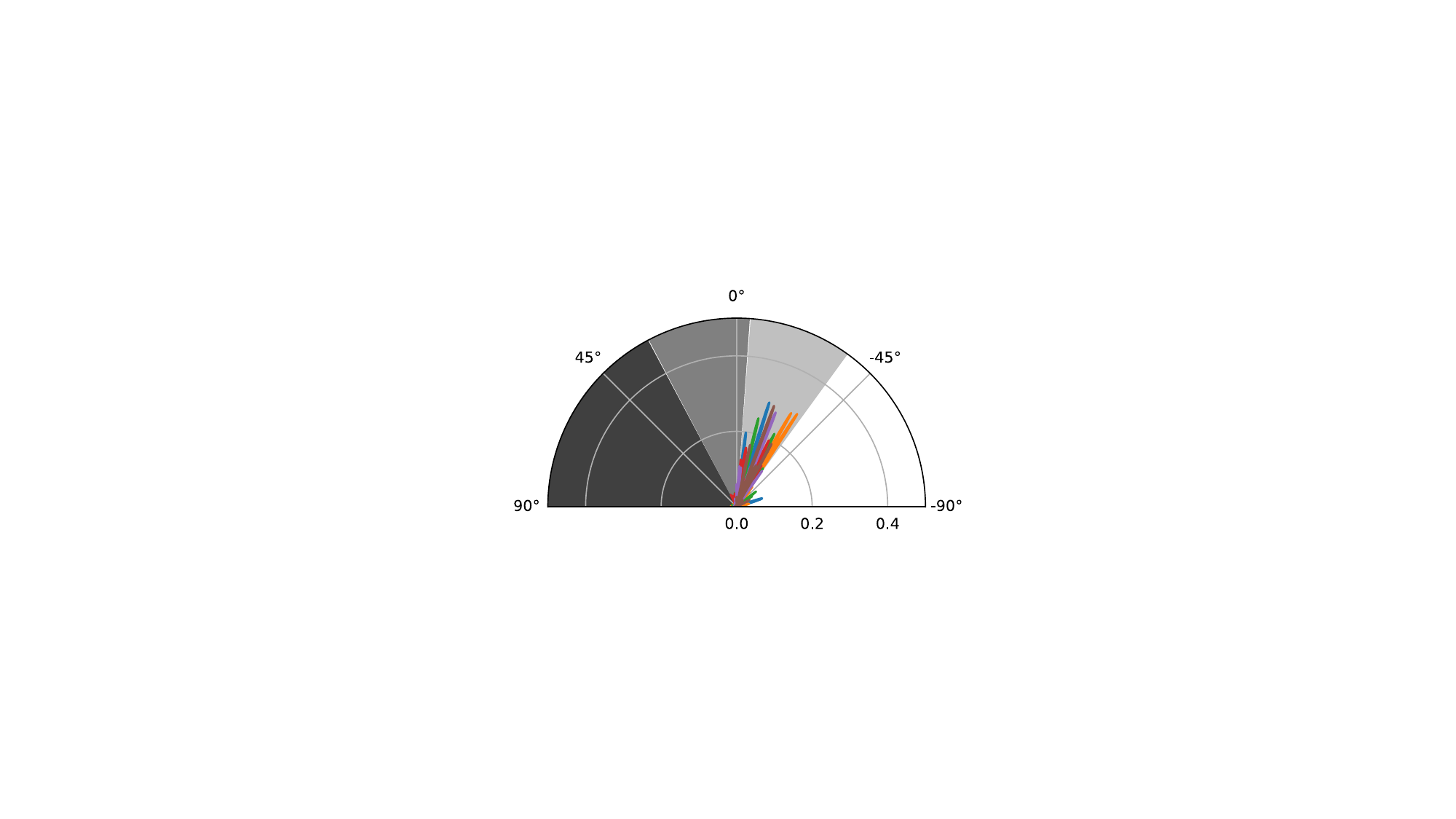}\label{fig6c}}
  \subfigure[$\mathbf{V}^{\rm{f}}_{3}$]
  {\includegraphics[width=.18\textwidth]{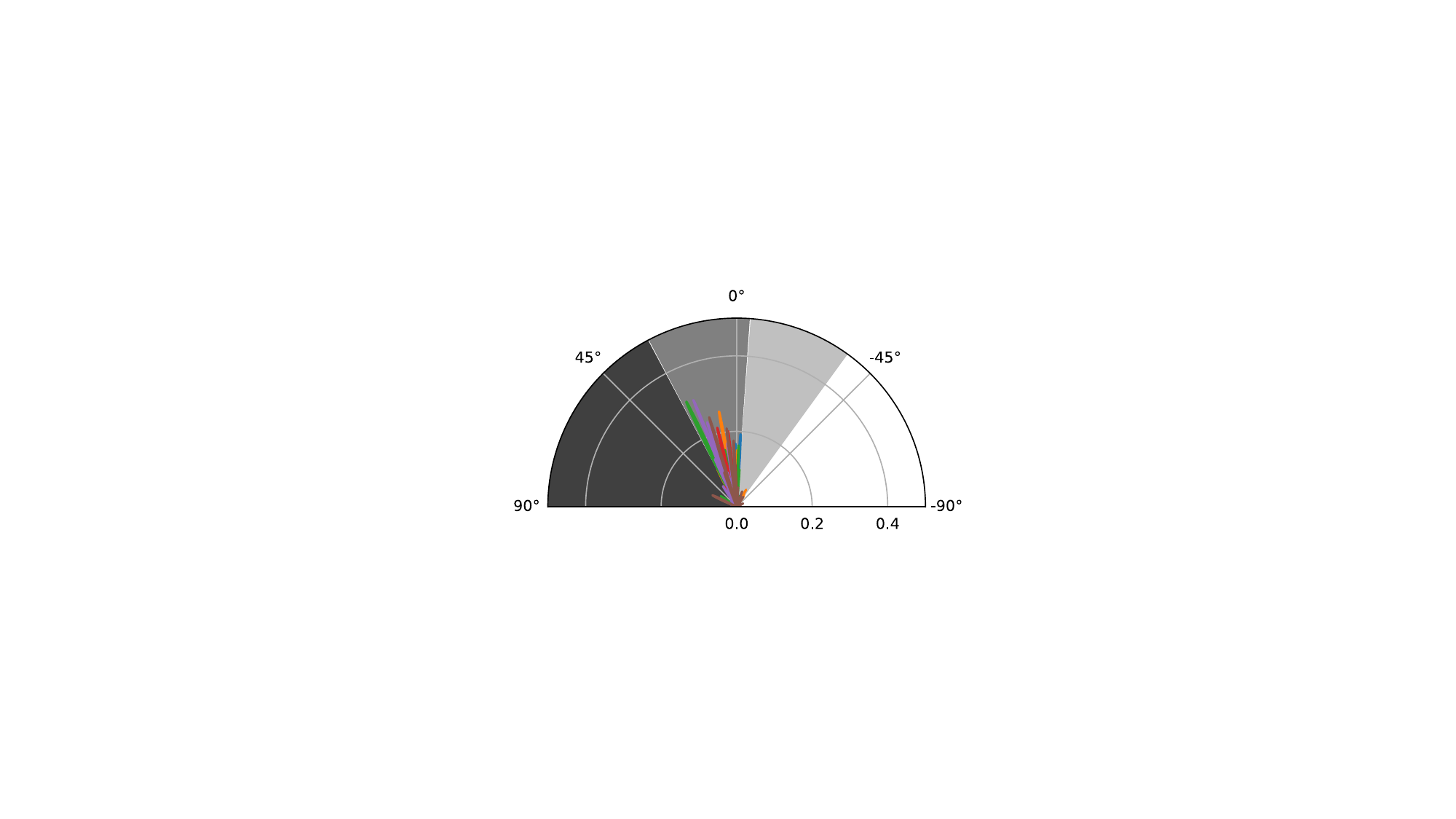}\label{fig6d}}
  \subfigure[$\mathbf{V}^{\rm{f}}_{4}$]
  {\includegraphics[width=.18\textwidth]{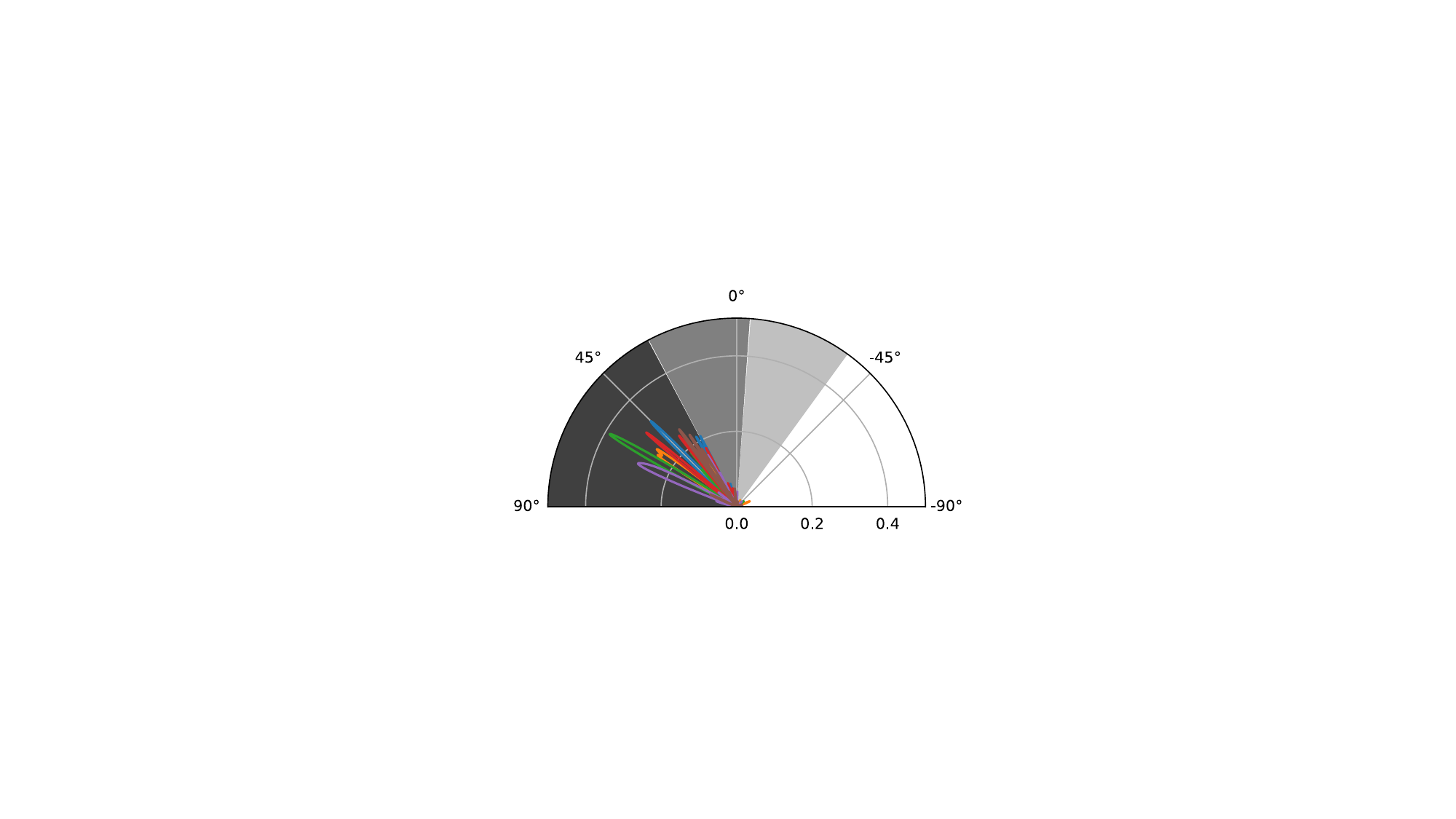}\label{fig6e}}

  \caption{The beam pattern of the learned PCs in DeepMIMO O1 for MISO systems.}\label{fig6}
  \vspace{-0.5cm}
\end{figure*}

\begin{figure*}[t]
  \centering
  \subfigure[$\mathbf{V}^{\rm{c}}$]
  {\includegraphics[width=.225\textwidth]{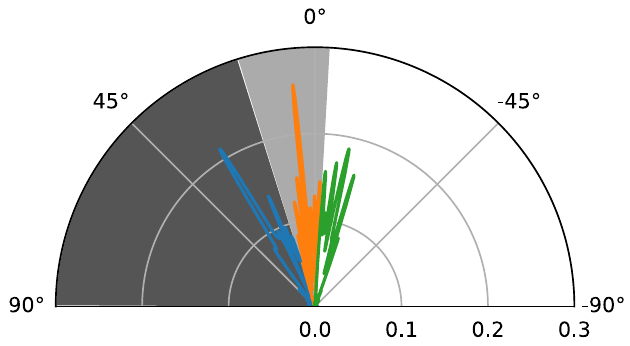}\label{fig7a}}
  \subfigure[$\mathbf{V}^{\rm{f}}_{1}$]
  {\includegraphics[width=.225\textwidth]{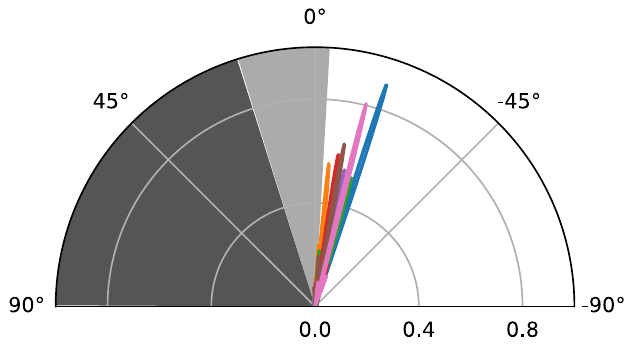}\label{fig7b}}
  \subfigure[$\mathbf{V}^{\rm{f}}_{2}$]
  {\includegraphics[width=.225\textwidth]{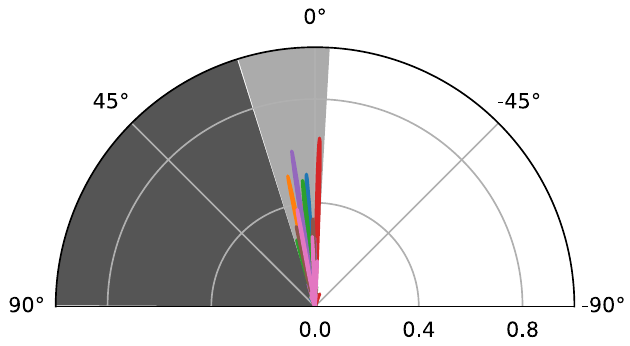}\label{fig7c}}
  \subfigure[$\mathbf{V}^{\rm{f}}_{3}$]
  {\includegraphics[width=.225\textwidth]{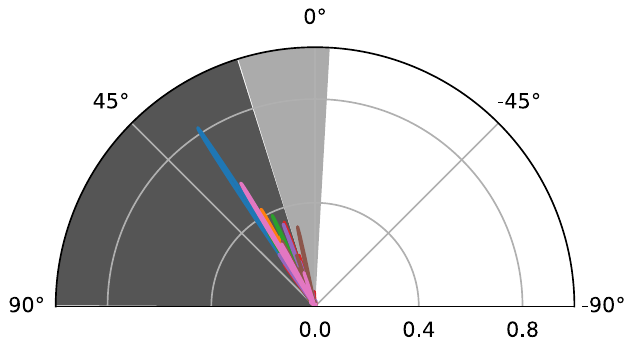}\label{fig7d}}

  \caption{The beam pattern of the learned PCs in DeepMIMO I3 for MISO systems.}\label{fig7}

\end{figure*}

\subsubsection{Learned PC patterns}
The two tiers of PCs are designed to collect channel information for the site-specific BS. Fig.~\ref{fig6} shows the PCs trained in the DeepMIMO O1 experiment, where Fig.~\ref{fig6a} is the pattern of the coarse-search codebook and the rest shows the fine-search codebooks. We set $N_1=4$ and $N_2=6$. Different grey levels are utilized to represent the groups generated by the K-means method, i.e., labels in the coarse-search part. It is worth noting that the pattern of coarse-search codebook $\mathbf{V}^{\rm{c}}$ exactly fits to the clustering results. Similar to the pattern of codebook $\mathbf{V}^{\rm{c}}$, each codebook in the fine-search part is learned to focus on a particularly narrower region. 
The PCs for the DeepMIMO I3 experiment are shown in Fig.~\ref{fig7}, with $N_1=3$ and $N_2=7$.
The UEs in the DeepMIMO I3 experiment are mainly located in front of the BS. Compared with the DeepMIMO O1 experiment, the beams in the DeepMIMO I3 experiment are much more focused. This indicates that the PC patterns can be trained to be adaptive for different environments.

\subsection{MIMO System}
Next, we consider the MIMO system where both the BS and the UE are equipped with multiple antennas. Here, we set $M_{\rm{t}} = 64$ and $M_{\rm{r}} = 16$. The adopted DFT codebooks $\mathbf{V}$ and $\mathbf{W}$ are set to contain $N_{\rm{t}} = 128$ and $N_{\rm{r}} = 32$ codewords, respectively. The transmit power in both experiments are set to $5$ dBm. 
The number of fine-search codebooks $G$ is also set as $G = 4$ for DeepMIMO O1 and $G = 3$ for DeepMIMO I3 since the changes in the number of antenna elements have no effect on UE distributions.

\subsubsection{Channel clustering}
\begin{figure}[t] 
\centering
\subfigure[DeepMIMO O1]
{\includegraphics[width=.40\textwidth]{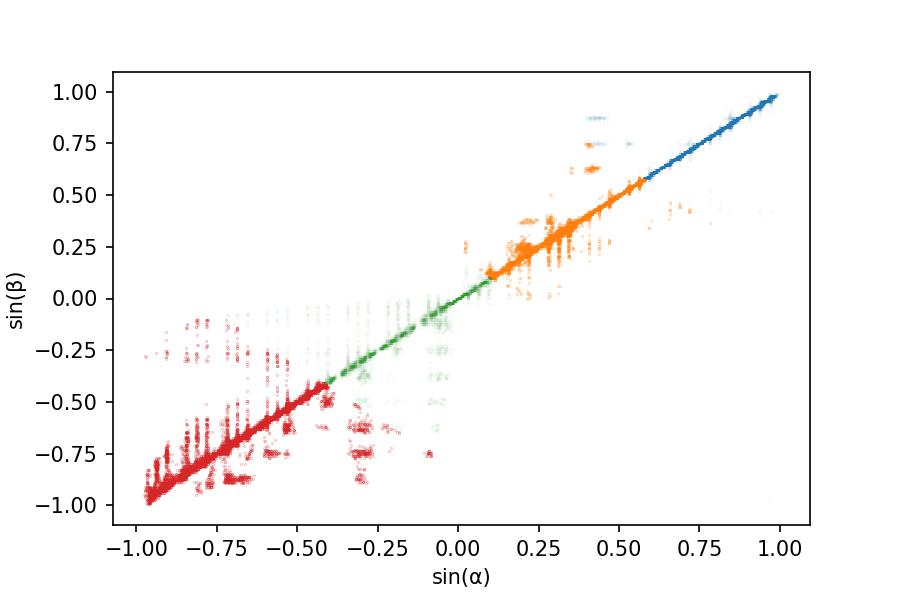}\label{k_means_a}}
\subfigure[DeepMIMO I3]
{\includegraphics[width=.40\textwidth]{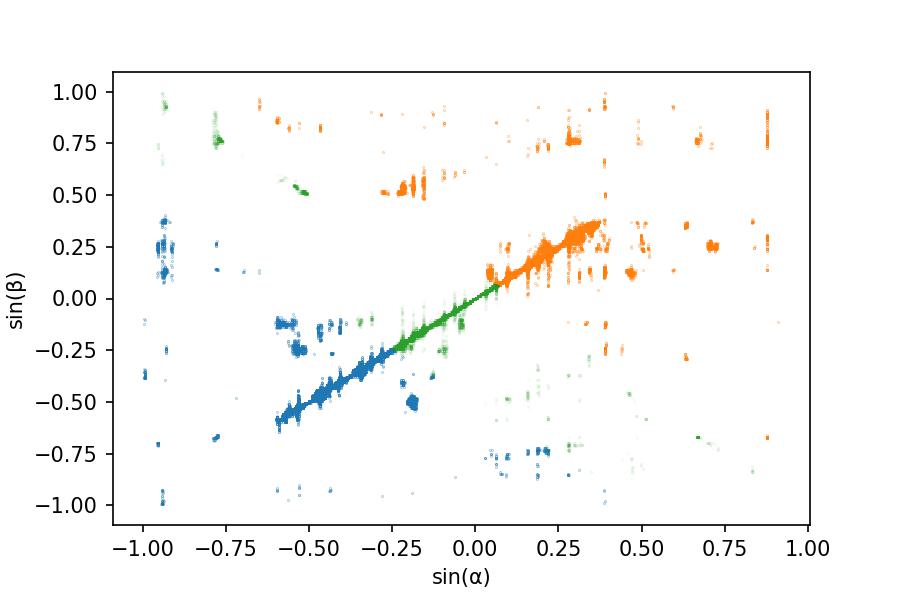}\label{k_means_b}}

\caption{Two-dimensional K-means clustering result of the UE distribution in MIMO systems.}\label{k_means_both}

\end{figure}

The proposed clustering method for channels in the MIMO system considers the optimal angle $\alpha$ and $\beta$ simultaneously for the beam alignment at both the BS and the UE. We show the result of the two-dimensional $K$-means clustering in Fig. \ref{k_means_both}. Here, each UE is marked as a scattered point with the coordinate $(sin(\alpha),sin(\beta))$ and different colors are utilized to differentiate clusters. For UEs which have LoS paths to the BS, the optimal discrete beam directions $\alpha$ at the BS and $\beta$ at the UE are supposed to be close. However, the NLoS UEs often only have reflecting links to the BS and thus each cluster may cover a wider region.  
Due to the large amount of the NLoS UEs in DeepMIMO I3, the optimal beam direction pairs of the UEs are distributed in almost the whole space.


\subsubsection{Performance evaluation}
In the scenario where both the BS and the UE perform beam alignment, the benchmarks are also modified for the MIMO system. Here, the definition of beam alignment accuracy is changed to be the probability of correctly predicting the optimal beams for both the BS and the UE.

\textbf{Separate HBAN-MISO:} The separate HBAN-MISO implements the HBAN-MISO at the BS and the UE, respectively. Specifically, the method utilizes independent PCs at the BS and the UE to collect channel information and predict the optimal beams sequentially. 


\textbf{One-tier PC based search:} 
This approach extends the DL-based methods in \cite{Jeffrey} to directly predicting the optimal beam pairs based on a learnable one-tier redesigned PC which contains a few probing beam pairs.


\textbf{Exhaustive search:}  
Since there are pre-defined DFT codebooks adopted at both the BS and the UE, the exhaustive search method sweeps all possible beam pairs and then selects the beam pair with the highest received power. The method can usually be regarded as a performance upper bound.

\textbf{Two-tier joint search:} There are two-tier AMCF codebooks adopted at both the BS and the UE. The method first sweeps all possible wide beam pairs to find the optimal wide beam at each side. Then all possible child beam pairs of the optimal wide beam are swept to find the optimal narrow beam pair.

\textbf{Two-tier hybrid search:} Similar to the two-tier joint search, the method first find the optimal wide AMCF beam pair by joint search. Then UE fixes its wide beam while the BS performs child beam sweeping to find the optimal narrow beam. Finally the BS fixes its narrow beam and the UE performs child beam sweeping to find the optimal narrow beam. The method is widely used in current 5G beam alignment framework \cite{5G_framework}.



\textbf{Binary joint search:} There are binary AMCF codebooks at both the BS and the UE. In each tier, the BS and the UE sweep all 4 possible beam pairs and find the best one, then determine the 4 beam pairs to sweep in the next tier. Since the binary codebook for the UE contains fewer tiers, the method first finds the optimal narrow beam of the UE. Then the UE fixes its narrow beam and the BS continues performing binary search to find its optimal narrow beam.

\begin{table}[t]
\newcommand{\tabincell}[2]{\begin{tabular}{@{}#1@{}}#2\end{tabular}}
\caption{Probing Codebook Sizes for HBAN-MIMO}\label{HBAN-MIMO}
\vspace{-0.4cm}
\begin{center}

\begin{tabular}{c|c c c c c c c c}

\hline
$N_1$ & 3 & 4 & 4 & 4 & 4 & 4 & 4 &  4\\
\hline
$N_2$ & 3 & 4 & 6 & 8 & 10 & 12 & 14 & 16\\
\hline
Sum & 6 & 8 & 10 & 12 & 14 & 16 & 18 & 20\\
\hline

\end{tabular}
\label{tab3}
\end{center}
\vspace{-0.4cm}
\end{table}

\begin{table}[t]
\newcommand{\tabincell}[2]{\begin{tabular}{@{}#1@{}}#2\end{tabular}}
\caption{Probing Codebook Sizes for Separate HBAN-MISO}\label{MIMO_benchmark}
\vspace{-0.4cm}
\begin{center}

\begin{tabular}{c|c c c c c c c c}

\hline
$N^{t}_1$ & 1 & 2 & 2 & 2 & 3 & 4 & 4 &  4\\
\hline
$N^{t}_2$ & 3 & 3 & 4 & 5 & 5 & 6 & 7 & 8\\
\hline
$N^{r}_1$ & 1 & 1 & 2 & 2 & 3 & 3 & 3 &  4\\
\hline
$N^{r}_2$ & 1 & 2 & 2 & 3 & 3 & 3 & 4 &  4\\
\hline
Sum & 6 & 8 & 10 & 12 & 14 & 16 & 18 & 20\\
\hline
\end{tabular}
\label{tab4}
\end{center}
\vspace{-0.4cm}
\end{table}

For HBAN-MIMO, the sizes of PCs in the coarse-search part and fine-search part for these two DeepMIMO experiments are shown in Table~\ref{HBAN-MIMO}. For separate HBAN-MISO, independent PCs are adopted at the BS and the UE. At the BS, we denote the sizes of PCs in the two tiers as $N^{t}_1$ and $N^{t}_2$. Similarly, sizes of PCs in these two tiers can be represented as $N^{r}_1$ and $N^{r}_2$ at the UE. Here, we show the above parameters for these two experiments in Table~\ref{MIMO_benchmark}. Finally, the size of the PC in the one-tier PC based search method is represented as $N_o$. All these parameters are determined based on simulation trials.

Here, we still measure the complexity of considered methods by the number of measurements, which is shown in Table~\ref{MIMO_complexity}. For methods with PCs, including HBAN-MIMO, separate HBAN-MISO, and one-tier PC based search, we adjust the above parameters properly and set $N_1+N_2 = N^{t}_1 + N^{t}_2+ N^{r}_1 + N^{r}_2= N_o$ in experiments for fair comparison. 
For the two-tier joint search method, we denote the sizes of wide-beam codebooks at the BS and the UE as $N_{\rm{W}}^{\rm{t}}$ and $N_{\rm{W}}^{\rm{r}}$, respectively. Then the number of measurements can be calculated as $N_{\rm{W}}^{\rm{t}}N_{\rm{W}}^{\rm{r}} + \lceil N_{\rm{t}}N_{\rm{r}}/(N_{\rm{W}}^{\rm{t}}N_{\rm{W}}^{\rm{r}})\rceil$. Note that the same notations can also be used for the two-tier hybrid search method, and the number of measurements for the method can be calculated as $N_{\rm{W}}^{\rm{t}}N_{\rm{W}}^{\rm{r}} + \lceil N_{\rm{t}}/N_{\rm{W}}^{\rm{t}} \rceil + \lceil N_{\rm{r}}/N_{\rm{W}}^{\rm{r}} \rceil$. To minimize the sweeping complexity of the two-tier joint search method, we set the sizes of wide-beam codebooks to be $N_{\rm{W}}^{\rm{t}}=16$ and $N_{\rm{W}}^{\rm{r}}=4$ so that the sizes of narrow-beam codebooks are $\lceil N_{\rm{t}}/N_{\rm{W}}^{\rm{t}} \rceil=8$ and $\lceil N_{\rm{r}}/N_{\rm{W}}^{\rm{r}}=8 \rceil$ at the two sides, respectively. We set the same parameters for the two-tier hybrid search method to make these two benchmarks adopt the same hierarchical codebook. The binary search method and the exhaustive search method have constant numbers of measurements if sizes of DFT codebooks $\mathbf{V}$ and $\mathbf{W}$ are determined. For the binary search method, there are $\log_2 N_{\rm{t}}$ tiers in the codebook for the BS and $\log_2 N_{\rm{r}}$ tiers in the codebook for the UE, where each tier consists of 2 codewords and thus the number of measurements is $4\log_{2}N_{\rm{r}} + 2(\log_{2}N_{\rm{t}}-\log_{2}N_{\rm{r}})$. For the exhaustive search method, there are totally $N_{\rm{t}}N_{\rm{r}}$ beam pairs for sweeping, which usually leads to a high complexity.

\begin{table}[t]
\newcommand{\tabincell}[2]{\begin{tabular}{@{}#1@{}}#2\end{tabular}}
\caption{Beam Sweeping Complexity of Different Methods in MIMO systems}\label{MIMO_complexity}
\vspace{-0.4cm}
\begin{center}

\begin{tabular}{c | c}

\hline
\hline
\textbf{Method} & \textbf{Complexity ($N_{\rm{t}}=128$, $N_{\rm{r}}=32$)}  \\
\hline
HBAN-MIMO & $N_1+N_2$  \\

Separate HBAN-MISO & $N^{t}_1 + N^{t}_2+ N^{r}_1 + N^{r}_2$ \\

One-tier PC based search & $N_o$ \\

Exhaustive search &  $128\times32 = 4096$ \\

Two-tier joint search &  $16\times4 + 8\times8 = 128$\\

Two-tier hybrid search & $16\times4 + 8+8 = 80$ \\

Binary search & $4\times\log_{2}{32} + 2\times(\log_{2}{128}-\log_{2}{32})=24$ \\

\hline
\hline
\end{tabular}
\end{center}

\end{table}

\begin{figure}[t] 
\centering
\subfigure[DeepMIMO O1]
{\includegraphics[width=.40\textwidth]{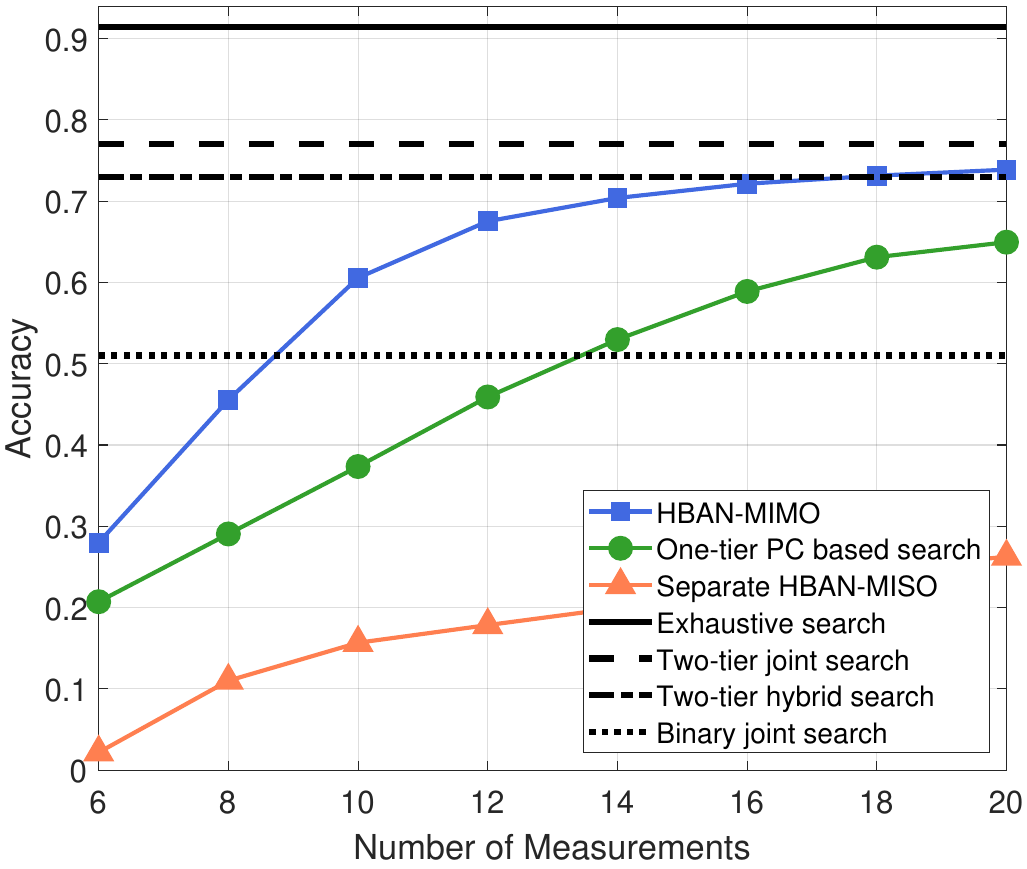}\label{deepmimo_O1_booksize_both}}
\subfigure[DeepMIMO I3]
{\includegraphics[width=.40\textwidth]{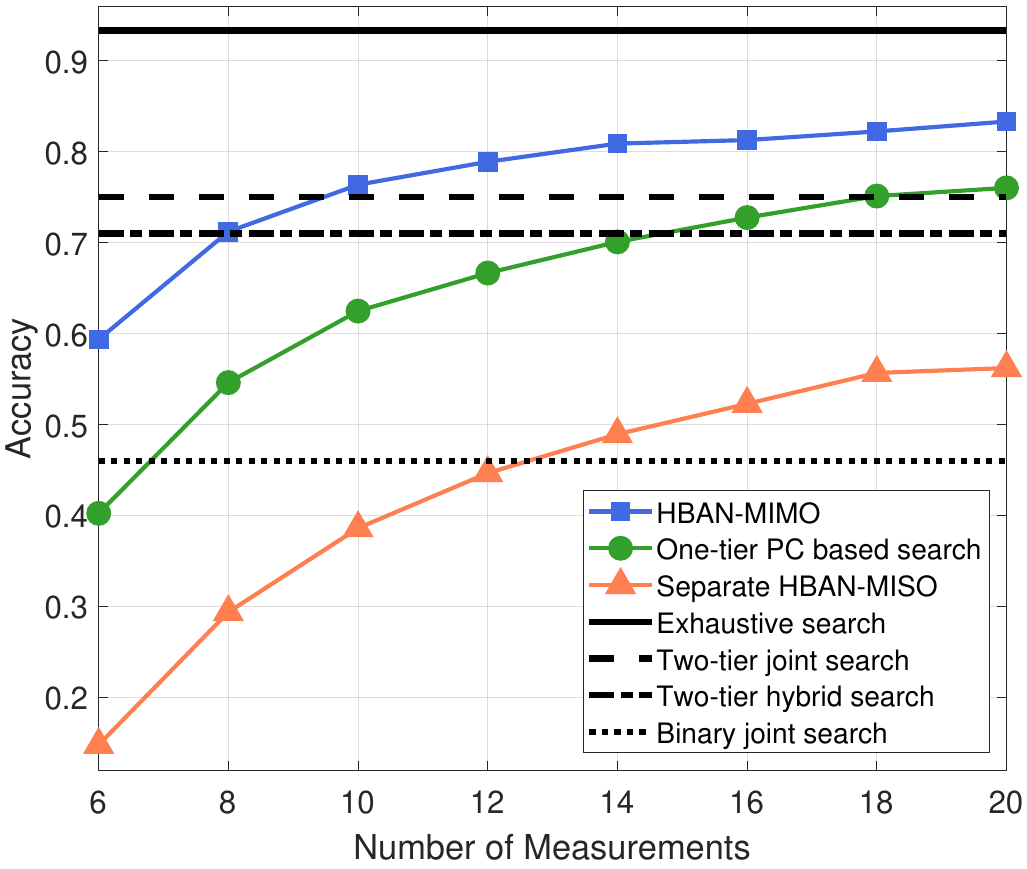}\label{deepmimo_I3_booksize_both}}

\caption{Beam alignment accuracy v.s. number of measurements in MIMO systems.}\label{booksize_both}

\end{figure}

\begin{figure}[t] 
\centering
\subfigure[DeepMIMO O1]
{\includegraphics[width=.40\textwidth]{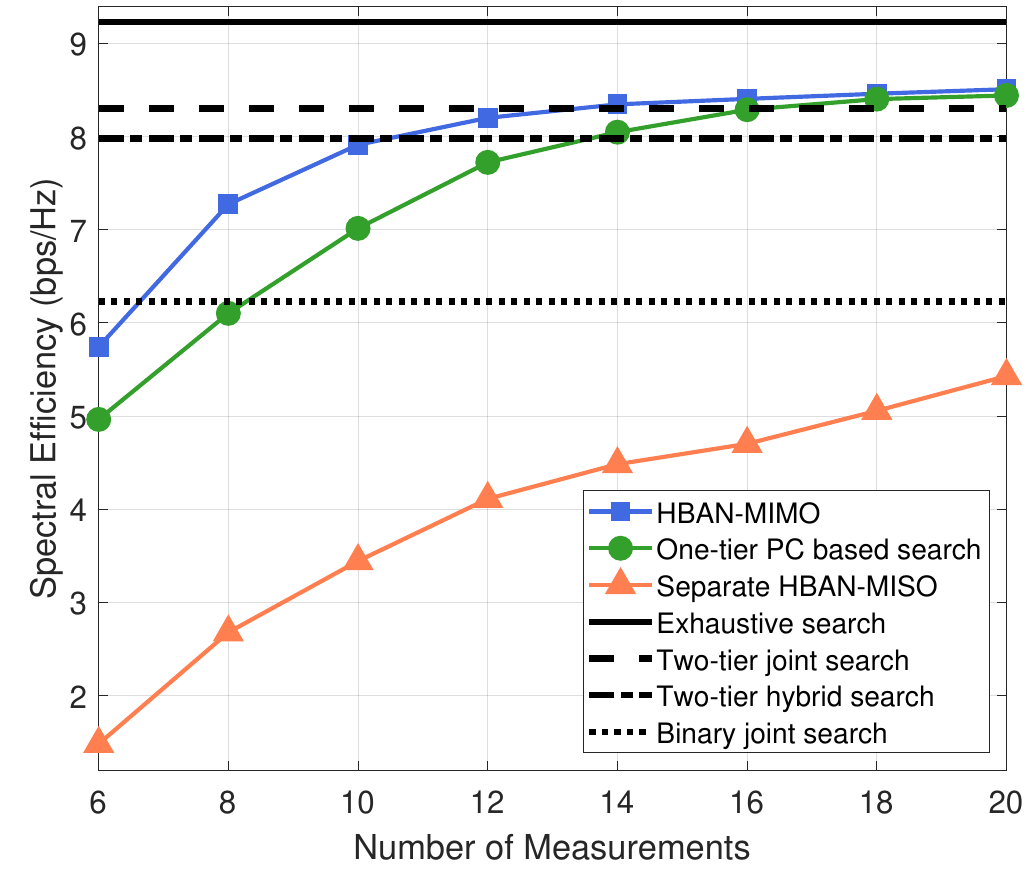}\label{deepmimo_O1_bps_both}}
\subfigure[DeepMIMO I3]
{\includegraphics[width=.40\textwidth]{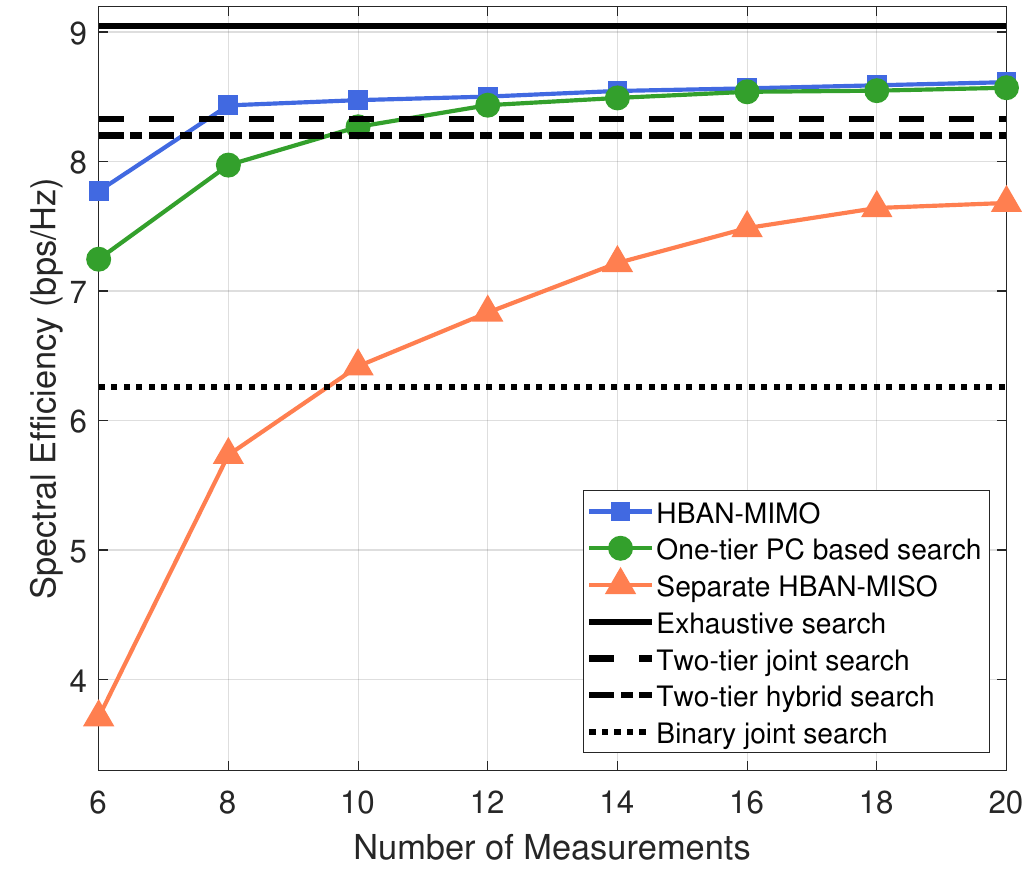}\label{deepmimo_I3_bps_both}}

\caption{Spectral efficiency v.s. number of measurements in MIMO systems.}\label{bps_both}

\end{figure}

The accuracy of considered methods with respect to the number of measurements is shown in Fig. \ref{booksize_both}. 
It is shown that HBAN-MIMO can consistently achieve a much better performance than separate HBAN-MISO, which suggests that performing beam alignment at the BS and the UE jointly can provide significant performance improvement. Furthermore, by comparing the performance of HBAN-MIMO and the one-tier PC based search method, we find that the hierarchical structure of HBAN-MIMO can assuredly bring about dramatic performance enhancement. 

Fig. \ref{bps_both} shows the spectral efficiency of considered methods under different numbers of measurements. We find that HBAN-MIMO outperforms all other benchmarks except the exhaustive search method with only 16 and 8 measurements in DeepMIMO O1 and DeepMIMO I3, respectively.
It is also observed that performance gaps between different methods in both experiments are much narrowed compared to Fig. \ref{booksize_both}. 
We can also find that even a small number of measurements can enable HBAN-MIMO to achieve a satisfactory performance. As such, the proposed method can save tremendous signaling overhead in practical systems.

\section{Conclusion}
In this paper, we investigate the hierarchical beam alignment method for the mmWave communication system. We propose a novel method which utilizes the measurements of hierarchical PCs to search for the optimal narrow beam in a coarse-to-fine way for the MISO system. A DNN-based architecture is designed to realize the learnable beamforming codebook, associated selector, and beam predictors. We also propose an effective two-step training strategy. Furthermore, we modify the proposed method by considering beam alignment in the MIMO system where both the BS and the UE are equipped with antenna arrays. Simulation results employing ray-tracing datasets show the superior performance of our method. Specifically, in the MISO system, our HBAN-MISO achieves a similar performance with much less beam sweeping overhead compared to the conventional two-tier hierarchical search and even outperforms the exhaustive search method since the learned two tiers of PCs can effectively capture the features of the propagation environment. In the MIMO system, the proposed HBAN-MIMO still achieves superior performance compared to representative existing alternatives by redesigning the codeword structure and performing beam alignment at the BS and the UE jointly. 
Future works may consider the effect of wideband beam squint on codebook design and beam prediction with an online training phase.

\bibliographystyle{IEEEtran}
\bibliography{refer_trans} 
	
\end{document}